\documentclass[lettersize,twocolumn]{IEEEtran}
\usepackage{amsmath,amsfonts,amssymb,caption,color,epsfig,graphics,graphicx,hyperref,latexsym,mathrsfs,revsymb,theorem,url,verbatim,epstopdf,mathtools,enumerate}
\usepackage{cite}
\usepackage{cellspace}
\usepackage{subfigure}\usepackage{makecell}\usepackage{multirow}\usepackage{diagbox}
\hypersetup{colorlinks,linkcolor={blue},citecolor={blue},urlcolor={blue}}

\usepackage[qm]{qcircuit}

\newtheorem{definition}{Definition}
\newtheorem{proposition}[definition]{Proposition}
\newtheorem{lemma}[definition]{Lemma}

\newtheorem{theorem}[definition]{Theorem}
\newtheorem{corollary}[definition]{Corollary}
\newtheorem{conjecture}[definition]{Conjecture}

\newtheorem{remark}[definition]{Remark}
\newtheorem{example}[definition]{Example}
\newtheorem{question}[definition]{Question}
\newtheorem{memo}[definition]{Memo}

\def\squareforqed{\hbox{\rlap{$\sqcap$}$\sqcup$}}
\def\qed{\ifmmode\squareforqed\else{\unskip\nobreak\hfil
\penalty50\hskip1em\null\nobreak\hfil\squareforqed
\parfillskip=0pt\finalhyphendemerits=0\endgraf}\fi}
\def\endenv{\ifmmode\;\else{\unskip\nobreak\hfil
\penalty50\hskip1em\null\nobreak\hfil\;
\parfillskip=0pt\finalhyphendemerits=0\endgraf}\fi}
\newenvironment{proof}{\noindent \textbf{{Proof.~} }}{\qed}
\def\Dbar{\leavevmode\lower.6ex\hbox to 0pt
{\hskip-.23ex\accent"16\hss}D}
\makeatletter
\def\url@leostyle{%
  \@ifundefined{selectfont}{\def\UrlFont{\sf}}{\def\UrlFont{\small\ttfamily}}}
\makeatother
\def\bcj{\begin{conjecture}}
\def\ecj{\end{conjecture}}
\def\bcr{\begin{corollary}}
\def\ecr{\end{corollary}}
\def\bd{\begin{definition}}
\def\ed{\end{definition}}
\def\bea{\begin{eqnarray}}
\def\eea{\end{eqnarray}}
\def\beq{\begin{equation}}
\def\eeq{\end{equation}}
\def\bal{\begin{aligned}}
\def\eal{\end{aligned}}
\def\bem{\begin{enumerate}}
\def\eem{\end{enumerate}}
\def\bex{\begin{example}}
\def\eex{\end{example}}
\def\bim{\begin{itemize}}
\def\eim{\end{itemize}}
\def\bl{\begin{lemma}}
\def\el{\end{lemma}}
\def\bma{\begin{bmatrix}}
\def\ema{\end{bmatrix}}
\def\bpf{\begin{proof}}
\def\epf{\end{proof}}
\def\bpp{\begin{proposition}}
\def\epp{\end{proposition}}
\def\bqu{\begin{question}}
\def\equ{\end{question}}
\def\br{\begin{remark}}
\def\er{\end{remark}}
\def\bt{\begin{theorem}}
\def\et{\end{theorem}}
\def\bmm{\begin{memo}}
\def\emm{\end{memo}}

\def\btb{\begin{tabular}}
\def\etb{\end{tabular}}

\newcommand{\nc}{\newcommand}

\def\a{\alpha}
\def\b{\beta}
\def\g{\gamma}

\def\e{\epsilon}
\def\ve{\varepsilon}

\def\t{\theta}

\def\l{\lambda}

\def\r{\rho}
\def\s{\sigma}

\def\ps{\psi}

\def\L{\Lambda}

\nc{\bbA}{\mathbb{A}} \nc{\bbB}{\mathbb{B}} \nc{\bbC}{\mathbb{C}}
 \nc{\bbD}{\mathbb{D}} \nc{\bbE}{\mathbb{E}} \nc{\bbF}{\mathbb{F}}
 \nc{\bbG}{\mathbb{G}} \nc{\bbH}{\mathbb{H}} \nc{\bbI}{\mathbb{I}}
 \nc{\bbJ}{\mathbb{J}} \nc{\bbK}{\mathbb{K}} \nc{\bbL}{\mathbb{L}}
 \nc{\bbM}{\mathbb{M}} \nc{\bbN}{\mathbb{N}} \nc{\bbO}{\mathbb{O}}
 \nc{\bbP}{\mathbb{P}} \nc{\bbQ}{\mathbb{Q}} \nc{\bbR}{\mathbb{R}}
 \nc{\bbS}{\mathbb{S}} \nc{\bbT}{\mathbb{T}} \nc{\bbU}{\mathbb{U}}
 \nc{\bbV}{\mathbb{V}} \nc{\bbW}{\mathbb{W}} \nc{\bbX}{\mathbb{X}}
 \nc{\bbZ}{\mathbb{Z}}

 \nc{\bA}{{\bf A}} \nc{\bB}{{\bf B}} \nc{\bC}{{\bf C}}
 \nc{\bD}{{\bf D}} \nc{\bE}{{\bf E}} \nc{\bF}{{\bf F}}
 \nc{\bG}{{\bf G}} \nc{\bH}{{\bf H}} \nc{\bI}{{\bf I}}
 \nc{\bJ}{{\bf J}} \nc{\bK}{{\bf K}} \nc{\bL}{{\bf L}}
 \nc{\bM}{{\bf M}} \nc{\bN}{{\bf N}} \nc{\bO}{{\bf O}}
 \nc{\bP}{{\bf P}} \nc{\bQ}{{\bf Q}} \nc{\bR}{{\bf R}}
 \nc{\bS}{{\bf S}} \nc{\bT}{{\bf T}} \nc{\bU}{{\bf U}}
 \nc{\bV}{{\bf V}} \nc{\bW}{{\bf W}} \nc{\bX}{{\bf X}}
 \nc{\bZ}{{\bf Z}}

\nc{\cA}{{\cal A}} \nc{\cB}{{\cal B}} \nc{\cC}{{\cal C}}
\nc{\cD}{{\cal D}} \nc{\cE}{{\cal E}} \nc{\cF}{{\cal F}}
\nc{\cG}{{\cal G}} \nc{\cH}{{\cal H}} \nc{\cI}{{\cal I}}
\nc{\cJ}{{\cal J}} \nc{\cK}{{\cal K}} \nc{\cL}{{\cal L}}
\nc{\cM}{{\cal M}} \nc{\cN}{{\cal N}} \nc{\cO}{{\cal O}}
\nc{\cP}{{\cal P}} \nc{\cQ}{{\cal Q}} \nc{\cR}{{\cal R}}
\nc{\cS}{{\cal S}} \nc{\cT}{{\cal T}} \nc{\cU}{{\cal U}}
\nc{\cV}{{\cal V}} \nc{\cW}{{\cal W}} \nc{\cX}{{\cal X}}
\nc{\cZ}{{\cal Z}}

\nc{\hA}{{\hat{A}}} \nc{\hB}{{\hat{B}}} \nc{\hC}{{\hat{C}}}
\nc{\hD}{{\hat{D}}} \nc{\hE}{{\hat{E}}} \nc{\hF}{{\hat{F}}}
\nc{\hG}{{\hat{G}}} \nc{\hH}{{\hat{H}}} \nc{\hI}{{\hat{I}}}
\nc{\hJ}{{\hat{J}}} \nc{\hK}{{\hat{K}}} \nc{\hL}{{\hat{L}}}
\nc{\hM}{{\hat{M}}} \nc{\hN}{{\hat{N}}} \nc{\hO}{{\hat{O}}}
\nc{\hP}{{\hat{P}}} \nc{\hR}{{\hat{R}}} \nc{\hS}{{\hat{S}}}
\nc{\hT}{{\hat{T}}} \nc{\hU}{{\hat{U}}} \nc{\hV}{{\hat{V}}}
\nc{\hW}{{\hat{W}}} \nc{\hX}{{\hat{X}}} \nc{\hZ}{{\hat{Z}}}

\nc{\hn}{{\hat{n}}}

\def\diag{\mathop{\rm diag}}
\def\dim{\mathop{\rm Dim}}

\def\min{\mathop{\rm min}}

\def\rank{\mathop{\rm rank}}

\def\supp{\mathop{\rm supp}}
\def\tr{\mathop{\rm Tr}}

\def\dg{\dagger}

\def\ox{\otimes}

\def\ra{\rightarrow}

\newcommand{\bra}[1]{\langle#1|}
\newcommand{\ket}[1]{|#1\rangle}
\newcommand{\proj}[1]{| #1\rangle\!\langle #1 |}
\newcommand{\ketbra}[2]{|#1\rangle\!\langle#2|}
\newcommand{\braket}[2]{\langle#1|#2\rangle}

\newcommand{\norm}[1]{\lVert#1\rVert}
\newcommand{\abs}[1]{|#1|}

\def\Dbar{\leavevmode\lower.6ex\hbox to 0pt
{\hskip-.23ex\accent"16\hss}D}
\begin{document}

\title{Mixed States Uniquely Determined by Marginals and Additivity}

\author{Xinyu Qiu, Lin Chen, Genwei Li, and Delin~Chu,~\IEEEmembership{Senior Member,~IEEE}
\thanks{Xinyu Qiu, Lin Chen, and Genwei Li are with LMIB (Beihang University), Ministry of Education, and School of Mathematical Sciences, Beihang University, Beijing 100191, China.}%
\thanks{Delin Chu is with the Department of Mathematics, National University of Singapore, Singapore 119076, Singapore.}%
\thanks{Corresponding author: Lin Chen (e-mail: linchen@buaa.edu.cn).}%
}

\maketitle
\begin{abstract}
Identifying whether a mixed quantum state is uniquely determined among all states (UDA) by its local marginals is a basic problem in quantum information theory. We establish necessary and sufficient conditions under which several classes of multipartite mixed states are UDA by their $k$-partite marginals. We also prove structural properties based on ranges and marginals, and formulate a recursive procedure for the determination of UDA states. We show that sufficiently high rank rules out unique determination from fixed-order marginals, implying that almost all multipartite mixed states are not UDA from such marginals. Finally, we completely characterize the additivity of bipartite UDA states, three-qubit and several families of $n$-qubit product UDA states.  These results clarify the boundary between UDA and non-UDA mixed states and provide a framework for local-marginal reconstruction and related certification tasks.
\end{abstract}


\section{Introduction}

Understanding the information of a global state with its local parts elucidates profound differences between quantum mechanics and its classical counterpart.   
Quantum marginal problem deals with
the existence of a valid global state, whose marginals (i.e., reduced density matrices) match the 
 input set of states \cite{coleman1963structure, klyachko2004quantum}. 
 If such a quantum marginal problem does admit a solution, one may focus on an interrelated problem.
Given access to local information  provided by the marginals, is  the global state determined uniquely by the marginals? If so, this global state is  uniquely determined among all (UDA) states.
This uniqueness is precisely the prerequisite for quantum state tomography via reduced density operators: only when the global state is uniquely determined by its marginals can the full state be reconstructed from local data alone \cite{bisio2009optimal,huang2025process, qt2017tao}. More broadly, this property is also relevant to quantum system certification, self-testing \cite{ivan2023quantum}, and other device-independent tasks \cite{qiu2024measurement}. 
Generally, the number of observables required for certification increases exponentially with the number of systems \cite{qin2024state}, whereas the UDA property allows the required number of observables to grow polynomially.
This  lowers the complexity of reconstructing the state of systems substantially \cite{cramer2010efficient,experiment2024hu}. Besides, the UDA property brings tools for dissipative quantum control \cite{karuvade2019uniquely}, and topological stabilizer codes in multipartite quantum error correction \cite{liao2022topo}. 
Moreover, the characterization of a set of marginals that uniquely determines the global state sheds light on the structure of multipartite entanglement across different subsystems \cite{den2002almost, walck2008only}. 

The UDA problem arises from the scope of pure states. 
For a pure state, one distinguishes between states uniquely determined among pure (UDP) states and states uniquely determined among all states \cite{qt2017tao}. 
The pure-state picture has been extensively studied.  Almost all three-qubit pure states, general symmetric Dicke and $W$-type states have been proved to be UDA by 2-marginals \cite{den2002almost, rana2011optimal, yu2013multi, zhang2024EvenParticle}.
The only $n$-particle pure states which cannot be UDP by their $(n-1)$-marginals are those GHZ-type states, and the result is further improved to the case of UDA \cite{walck2008only}. These results show that uniqueness from local information is often generic for pure states. 
The mixed-state case is substantially different and is also more relevant experimentally, since realistic quantum systems inevitably interact with their environments and lose purity through noise and decoherence \cite{liu2023observation, lee2025topo}. Despite this practical importance, mixed states uniquely determined by local marginals remain much less understood. Existing results mainly establish uniqueness for almost all states in certain regimes or for specific low-rank families \cite{role2014chen, uinque2013chen}. Although such results are valuable, they do not directly decide whether a given concrete state is UDA, because the exceptional measure-zero set or the hidden structural assumptions may be difficult to verify \cite{jones2005parts, wyder2017almost, parashar2009nqubit}. This motivates the first goal of this work, i.e. to derive  necessary and sufficient conditions for mixed UDA states in concrete multipartite families.
 
For a set of marginals on subsystems of fixed size, the global state may transition from a UDA state to a non-UDA state as the number of systems increases. For example, generic states of $n$-qudits are UDA by certain sets of marginals of less than about two-thirds \cite{linden2002parts}, or
more than half of the parties \cite{jones2005parts}, whereas the reduced states of fewer than half of the parties are not sufficient. 
 This implies that given two mixed UDA states $\a$ and $\b$, their additive form $\a\ox \b$ is not necessarily UDA. Admitting that one of $\a$ or $\b$ is pure, $\a\ox \b$ is shown to be UDA by three composite types of tensor product \cite{add2023shen}.  
Generally, it is worth considering the additivity of two mixed UDA states by focusing on the following questions: can the UDA additivity hold without additional conditions? If the answer is negative,  which kind of additional conditions of $\a$ and $\b$ make the UDA additivity hold?
The purity assumption in \cite{add2023shen} avoids analyzing the essence of UDA states, whereas this is unavoidable when both states are mixed. By fully characterizing mixed UDA states, the additivity of these UDA states can be further studied. This is the second motivation of this work. 

In this work, we study the complete characterization of several classes of multipartite mixed UDA states, establish the procedure for the determination of additional UDA states by the characterization, and show the necessary and sufficient condition under which the additivity of such UDA states holds. 
To begin with, we introduce some essential properties for arbitrary UDA states in Lemmas \ref{le:lr+(1-l)s_not2uda} and \ref{le:general_properties}. 
In detail, we show that the global state is not UDA if its marginals or faces of the range are not UDA, and that two states with the same range, or up to permutations and reversible channels, are either UDA or not at the same time. We also show the range inclusion between a specific state and its compatible states.  By these facts, we show the complete characterization of bipartite 1-UDA states $\r_{A_1A_2}$ in Proposition \ref{pro:bipartite1-UDA}. The necessary and sufficient conditions under which  the three-qubit state $\r_{A_1}\ox \r_{A_2A_3}$ in Theorem \ref{th:rhoA1 ox A23 iff}, four-qubit state $\r_{A_1A_2}\ox \r_{A_3A_4}$ in Theorem \ref{th:RA12oxRA34}, five-qubit state $\r_{A_1}\ox \r_{A_2A_3}\ox \r_{A_4A_5}$ in Theorem \ref{th:5-qubit} are 2-UDA, and three types of product $n$-qubit states in Proposition \ref{pro:n-qubit} are $k$-UDA. 
We clarify that the study of $k$-UDA, particularly for small $k$, is motivated by practical considerations. Since the measurement cost, the error of gates and the impact of noise  grow with $k$, minimizing resource consumption in local measurements makes $2$-UDA a favorable scenario.  
We also show that some mixed states with sufficiently high rank are definitely not UDA even with large-body marginals in Proposition \ref{pro:rankn-1}. It implies that almost all mixed states are not UDA, which is different in pure-state scenarios.  
Inspired by the proof of preceding facts, a systematic procedure is established for determining additional UDA states in FIG. \ref{fig:k-UDA}, and hence the determination is not restricted to product mixed states characterized earlier. 
The UDA additivity of bipartite 1-UDA, three-qubit product 2-UDA and some $n$-partite $k$-UDA states is further analyzed.  In detail, for the bipartite UDA states $\a$ and $\b$, $\a\ox\b$ is UDA if and only if one of $\a$ and $\b$ is a pure product state, shown in Theorem \ref{th:add2-partite}. The additivity  of three-qubit product 2-UDA states is considered in Theorem \ref{th:add3-qubit} by showing the necessary and sufficient condition, and that of $n$-qubit product states is analyzed subsequently. These results show that additivity is exceptional rather than automatic: it holds only under additional structural restrictions, typically involving purity or rigid low-rank blocks. The applications of mixed UDA states and their characterization are shown from the perspectives of related certification tasks.

The rest of this paper is organized as follows. In Sec. \ref{sec:preliminaries}, we clarify some notations and introduce the definitions and known facts. In Sec. \ref{sec:mixedUDA}, we characterize multipartite mixed UDA states and establish a systematic method to determine some non-$k$-UDA states. In Sec. \ref{sec:additivity}, we analyze the additivity of mixed UDA states and present the necessary and sufficient conditions. We show the applications of mixed UDA states and their characterization in Sec. \ref{sec:application}.  We conclude in Sec. \ref{sec:conclusion}.

\section{Preliminaries}
\label{sec:preliminaries}

In this section, we present the notation and the definition of mixed UDA states, along with some known facts used later.
    First, we introduce some notations. Suppose $A_1, A_2, ..., A_n$ are $n$ systems associated with the Hilbert spaces $\cH_{A_1}, \cH_{A_2}, ..., \cH_{A_n}$, respectively. We denote a state $\r$ on $\cH_{A_1}\ox \cH_{A_2}\ox  ...\ox \cH_{A_n}$ by $\r_{A_1A_2...A_n}$. We denote the range and kernel of the state $\r$ by $\cR(\r)$ and $\ker(\r)$, respectively. They are the range and kernel of the Hermitian density operator of $\r$. We have $\cR(\r)=\supp \r=(\ker(\r))^\perp$. Let $\cS$ be a subset of $\{A_1,A_2,...,A_n\}$ with $\abs{\cS}=k$ and $\cS^c=\{A_1,A_2,...,A_n\}\backslash \cS$. A $k$-partite marginal of $\r$ is given by $\r_{\cS}=\tr _{\cS^c}\r$. A state $\s$ is said to be  $k$-compatible with $\r$ if $\s$ has the same $k$-partite marginals as $\r$, i.e. $\s_{\cS}=\r_{\cS}$ for any $\cS$.  

Next we introduce the definition of a state
uniquely determined by its $k$-partite marginals. For a pure state $\ket\ps$, if there is no other pure state $k$-compatible with $\ket\ps$, then $\ket\ps$ is said to be $k$-partite uniquely determined among  pure states ($k$-UDP); if there is no other (pure or mixed) state $k$-compatible with $\ket\ps$, then $\ket\ps$ is said to be $k$-partite uniquely determined among all ($k$-UDA)  states.   This definition can be generalized to the case of mixed states by assuming that the initial state $\r$ is mixed. 
\begin{definition}
Let $\r$ be an arbitrary state (pure or mixed). If there is no other state having all the same $k$-partite marginals as $\r$, then $\r$ is called $k$-uniquely determined among all ($k$-UDA) states. 
\end{definition}
The existence of mixed $k$-UDA states is established in previous work. It is known that 
some mixed states with low rank can be UDA. For example, Theorem 4 in \cite{uinque2013chen}
shows that almost every tripartite density operator $\r$ acting on the Hilbert space $\cH_{d_1} \ox \cH_{d_2} \ox \cH_{d_3}$ with rank no more than $\lfloor\frac{d_1}{d_3}\rfloor$ can be uniquely determined among all states by its 2-marginals of partial pairs $\{1,2\}$ and $\{1,3\}$.
It is also shown in \cite{uinque2013chen} that giving a complete characterization of these undetermined tripartite states is not easy, as the methods proposed before do not directly provide it.  This inspires us to give a complete characterization of UDA states before analyzing their additivity. 
That is, we establish a criterion to determine whether a mixed state is UDA directly, which is exactly what we will show in Sec. \ref{sec:mixedUDA}. 

 A state with fewer parties may transform from a $k$-UDA state into a non-$k$-UDA one, with the increase of the number of parties with a fixed $k\in \bbN$. A construction of $k$-UDA states in a system with more parties is considered in \cite{add2023shen}.  In detail, three composite types of tensor product for two $k$-UDA states preserve $k$-UDA if one of the two $k$-UDA states is pure. This relies on the following fact, which we shall use later. 
 \begin{lemma} 
 	\label{le:pure sep} 
 	 If the reduction of the system $(A_1, \ldots, A_m)$ from the global state $\rho_{A_1 \cdots A_m E}$ is a pure state, then the global state is in the form
 	\begin{eqnarray}
 		\label{eq:r=ps ox s}
 		 	\rho_{A_1 \cdots A_m E}=\proj{\ps}_{A_1 \cdots A_m} \otimes \sigma_E.
 	\end{eqnarray}
 \end{lemma}
 Obviously, Lemma \ref{le:pure sep} does not generally hold  if the $m$-partite marginal $\rho_{A_1 \cdots A_m}$ is a mixed state. For example, the state $\r_{ABC}=p\proj{000}+(1-p)\proj{W}$ for $0<p<1$ and $\ket{W}=\frac{1}{\sqrt{3}}(\ket{100}+\ket{010}+\ket{001})$ satisfies that $\r_{AB}$ and $\r_C$ are both mixed, and $\r_{ABC}\neq \r_{AB}\ox \r_{C}$. Hence, the construction  in \cite{add2023shen} fails if both initial states are mixed. It is interesting to establish additional conditions on the initial states, by which the additivity of mixed UDA states still holds.    

We introduce some basic properties for a $k$-UDA state  as follows \cite{add2023shen}.
Let $\r$ be an $n$-partite $k$-UDA state. Then 
\begin{itemize}
    \item any state that is LU equivalent to $\r$ is also a $k$-UDA state;
    \item  $\r$ is also $(k+1)$-UDA state for $k+1\leq n$, and the converse does not generally hold. 
\end{itemize}
These facts allow us to simplify the form  of  density operators in the analysis of their UDA properties.

\section{The characterization of  multipartite mixed UDA states}
\label{sec:mixedUDA}
In this section, we show the necessary and sufficient conditions with which the bipartite to $n$-partite mixed state is UDA, and establish a systematic method to determine some non-$k$-UDA states. 
We present the essential properties of UDA states in Sec.  \ref{sec:preLemma}. By these properties, we show the necessary and sufficient condition by which a bipartite mixed state is 1-UDA in Sec. \ref{sec:bipartite_state}. We further characterize the tripartite  and four-qubit mixed 2-UDA states in Sec. \ref{sec:tripartite_state} and \ref{sec:four-qubit}, respectively. We analyze the $n$-qubit mixed $k$-UDA states for $n\geq 5$, and establish a systematic method to determine some non-$k$-UDA states in Sec. \ref{sec:n-qubit}. Our complete characterization of tripartite to $n$-partite states mostly focus on 2-UDA states, as such states require less local information than $k$-UDA states for $k\geq 3$. Hence, 2-UDA  states are of practical interest. 

\subsection{General properties of $k$-UDA states}
\label{sec:preLemma}
We show some general properties of $k$-UDA states from the perspective of their marginal compatibility, range structure and local channel invariance. In particular, Lemma \ref{le:lr+(1-l)s_not2uda} gives several basic closure and invariance properties of the $k$-UDA property, including its behavior under tensor products, convex mixtures, range equivalence, and subsystem permutations. Lemma \ref{le:general_properties} further characterizes the structure of $k$-compatible states through their real and imaginary parts, range inclusion relations, and local channel transformations. These results provide a structural framework for analyzing $k$-UDA states and will serve as basic tools in the subsequent discussions. The proof of Lemmas \ref{le:lr+(1-l)s_not2uda} and \ref{le:general_properties} is given in Appendix \ref{app:lemma}.

For any subspace $E\subseteq \bbC^m\ox \bbC^n$, the face of $E$ is defined as the
set of all states whose range is contained in $E$.
The following facts show that a state  with non-$k$-UDA tensor marginals, or the face of $\cR(\rho)$ has a non-$k$-UDA state then $\rho$ is not $k$-UDA. It is also shown that two states with the same range, or up to a system permutation are $k$-UDA simultaneously, and that two global states are $k$-compatible, implying that their marginals are also $k$-compatible. 
\begin{lemma}
\label{le:lr+(1-l)s_not2uda}
 Let $\r$ and $\s$ be two $n$-partite states. Then 

(i) If one of the states $\r$ and $\s$ is not $k$-UDA. Then $\r\ox \s$ is also not $k$-UDA.
 
 (ii) If $\r$ is not $k$-UDA, the convex combination of $\r$ and $\s$, i.e. $\l \r+(1-\l)\s$ is also not $k$-UDA, for $\l\in (0,1]$. Hence, if the face of $\cR(\rho)$ has a non-$k$-UDA state, then $\rho$ is not $k$-UDA.

 (iii) If $\r$ and $\s$ have the same range, then they are at the same time k-UDA or not.   

(iv) Two states $\r\ox \s$ and $\s\ox \r$ are the same time $k$-UDA or not.

 (v) Suppose $\r$ and $\s$ are on $\cH_{A_1A_2...A_n}$ are $k$-compatible with $n>k$. Let $\cS \subseteq (A_1,...,A_n)$ such that $|\cS| \geq k$. Then $\r_\cS$ and $\s_\cS$ are also $k$-compatible.
\end{lemma}

Lemma \ref{le:lr+(1-l)s_not2uda} naturally implies the following results.  First, the "additivity" to be analyzed can only be narrowed within the scope of two $k$-UDA states. Second, the systematical construction of a series of $k$-UDA
and non-$k$-UDA states is established. In detail, any state whose range is included in a $k$-UDA state is also $k$-UDA, and 
any mixed state whose convex decomposition includes a non-$k$-UDA state is also not $k$-UDA.  Apart from the simplification under LU equivalence, we are allowed to simplify the expression of states by keeping their range invariant. If $\a$ is not $k$-UDA and $\cR(\a)\subseteq \cR(\b)$, then $\b-\e \a $ is also not $k$-UDA, for $\e>0$ small enough.  Generally, the UDA property can only be invariant up to LU equivalence but the nonlocal operation,   permutation, can also preserve the UDA property of states. In this paper, we shall interchange the order of any two subsystems of a product state without explicit clarification.

We present several properties of 
$k$-UDA states and of pairs of 
$k$-compatible states. To characterize how a given state is related to all states compatible with it, it is useful to study their relations through the ranges of the corresponding density operators.  Moreover, certain local quantum channels preserve 
$k$-compatibility, and reversible channels even preserve the 
$k$-UDA property. 
\begin{lemma}
\label{le:general_properties}
(i) If two multipartite states $\r$ and $\s$ are $k$-compatible, then their real and imaginary parts are also $k$-compatible.    

(ii) Let $\r=\text{Re}(\r)+i \text{Im}(\r)$. Then $\cR(\text{Im}(\r))\subseteq \cR(\text{Re}(\r))$.
In particular, if $\text{Im}(\r)\neq 0$ and all the $k$-marginals of $\r$ are real, then $\r$ is not $k$-UDA. 

(iii) Suppose the  $n$-partite state $\rho\geq 0$ and $H$ is an Hermitian matrix of the same size, and $\cR(\rho) \supseteq \cR(H)$. Then $\cR(\rho_\cS) \supseteq \cR(H_\cS)$ for $\cS\subseteq[A_1A_2...A_n]$. The converse does not generally hold.

(iv) Suppose $\a$ and $\b$ are $k$-compatible $n$-partite states on systems $[A_1A_2...A_n]$, and $\L$ a channel on the first system $A_1$, with the remaining systems denoted as $B:=A_2...A_n$. Then (a) $(\L_{A_1} \otimes I_B) \alpha$ and $(\L_{A_1} \otimes I_B) \b$ are also $k$-compatible;
(b) if $\L$ is a reversible channel, then $\alpha$ is $k$-UDA iff $(\L_{A_1} \otimes I_B) \a$ is $k$-UDA. 

(v) Let $\r=\r_{A_1A_2...A_n} \otimes \r_{A_{n+1}A_{n+2}...A_{m}}$ be a $m$-partite state for $m\geq 2n$. Suppose $\r_{A_{n+1}A_{n+2}...A_{m}}$ is $n$-UDA. Then the $m$-partite state $\sigma$ $n$-compatible with $\r$ satisfies $\cR(\sigma) \subseteq \cR(\rho)$. 
\end{lemma}

Lemma \ref{le:general_properties} provides a structural understanding of $k$-UDA property. Intuitively, Lemma \ref{le:general_properties} (i) and (ii) show that $k$-compatibility is governed by linear constraints coming from the $k$-marginals, so it is inherited by the real and imaginary parts of a state.
Lemma \ref{le:general_properties} (iii) shows that global range inclusion descends to all reduced states, so range inclusions yield  necessary conditions at the level of marginals, even though the converse generally fails. Lemma \ref{le:general_properties} (iv) shows that both $k$-compatibility and, under reversible local channels, $k$-UDA are stable under local operations. Finally, Lemma \ref{le:general_properties} (v) shows that attaching an $n$-UDA factor can  restrict the freedom of other $n$-compatible states by forcing them to lie inside the same global range. 
If $\s$ is compatible with $\r$, the operator $\chi:=\r-\s$ for any $\s$ satisfies that any $n$-marginals are zero. One can verify that $\r$ is $n$-UDA if and only if $\chi=0$ for any $\s$. Such a restriction derives $\cR(\chi)\subseteq \cR(\r)$, by which we show that $\chi=0$ for some $\r$.

\subsection{On bipartite mixed UDA states}
\label{sec:bipartite_state}
We show the necessary and sufficient condition by which a bipartite mixed state is 1-UDA. The additivity of 1-UDA states $\a$ and $\b$ is explored by this characterization. 
\begin{proposition}
	\label{pro:bipartite1-UDA}
	A bipartite mixed state $\r$ on systems $A_1$ and $A_2$ is 1-UDA if and only if one of $\r_{A_1}$ and $\r_{A_2}$ is pure.  Hence if $\r_{A_2A_3}$ is 1-UDA then $\r=\r_{A_1}\ox \r_{A_2A_3}$ is 2-UDA.
\end{proposition}
\begin{proof}
	The if part is trivial by Lemma \ref{le:pure sep}. Next, we show the "only if" part. Let $\g_{A_1A_2}$ be a bipartite mixed state. The UDA property is invariant up to local unitaries. It suffices to discuss the UDA property of $\r_{A_1A_2}=(U\ox V) \g_{A_1A_2}(U^\dg \ox V^\dg)$. Note that 
	\begin{eqnarray}
		\label{eq:LUrhoA12}
		\r_{A_1}=U \g_{A_1} U^\dg, \quad
		\r_{A_2}=V \g_{A_2} V^\dg.
	\end{eqnarray}
	We can choose appropriate local unitaries $U$ and $V$, such that $\r_{A_1}$ and $\r_{A_2}$ are both diagonal.
	Let $\dim(\cH_{A_k})=d_{k}$ for $k=1,2$. Then $\r_{A_1A_2}$ can be given by 
$		\r_{A_1A_2}=\sum_{i,j=1}^{d_{1}}\ketbra{i}{j}\ox M_{ij}$, where $M_{ij}\in \bbC^{d_{2}\times d_{2}}$, and 
\begin{eqnarray}
	\label{eq:rho12}
	\r_{A_1}=\sum_{i,j=1}^{d_{1}} \ketbra{i}{j}\ox \tr M_{ij},
\quad
	\r_{A_2}=\sum_{j=1}^{d_{1}}  M_{jj},
\end{eqnarray}
 are both diagonal states. We further assume that $\r_{A_1A_2}=(r_{jk})$ for $j,k=1,2,...,d_{1}d_{2}$.  
	If there is an element $r_{jk}\neq 0$ for $j\neq k$, then $		\r_{A_1A_2}$ is not 1-UDA. In fact, we can choose the state $\s'_{A_1A_2}$ whose diagonal elements are the same as those of $\r$, and off-diagonal elements are half of those of $\r$, that is,  
	\begin{eqnarray}
		\s'=\bma r_{11} & \frac{r_{12} }{2}&  \cdots &  \frac{r_{1, d_{1}d_{2}}}{2}\\
		\frac{r_{12}^* }{2}& r_{22} & \cdots & \frac{r_{2,d_1d_2}}{2}\\
		\vdots & \vdots &  & \vdots\\
		\frac{r_{1,d_1d_2}^* }{2}& \frac{r_{2,d_1d_2}^*}{2} & \cdots & r_{d_1d_2,d_1d_2} \ema.
	\end{eqnarray}
	We have $\s'\neq \r$ and $\s' \geq 0$ by $\s'- \frac{1}{2}\r\geq 0$.
	From (\ref{eq:rho12}), one can verify that $\s'_{A_k}=\r_{A_k}$, i.e., $\s'$ is compatible with $\r$ by two 1-marginals.  Hence,  a 1-UDA bipartite state should be diagonal. 
	
	Next we show that $\r_{A_1A_2}$ is 1-UDA implies that 
 $\r_{A_1}$ or $\r_{A_2}$ is a pure state. Otherwise it holds that $\rank(\r_{A_k})\geq 2$. Since $\r_{A_k}$'s are diagonal, there is another state 
 $\s_1= \eta \ox \xi$ that is compatible with $\r$, where $\eta=\r_{A_1}:=\diag\{a_1,a_2, ...,a_{d_1}\}$ and 
 $\xi=\r_{A_2} :=\diag\{b_1,b_2, ...,b_{d_2}\}$.
Up to local unitaries, we assume that $a_1\geq a_2>0$ and 
$b_1\geq b_2>0$. There is a $0<\ve<\min\{a_1,a_2,b_1,b_2 \}$.   The state $\s_1$ can be decomposed as 
\begin{eqnarray}
	\s_1=(\a_1+\a_2)\ox (\b_1+\b_2),
\end{eqnarray}
	where $\a_1=\b_1=\diag\{\ve,\ve,0,...,0\}$, $\a_2=\diag\{a_1-\ve, a_2-\ve, a_3, ... , a_{d_1}\}$
	and $\b_2=\diag\{b_1-\ve, b_2-\ve, b_3, ... , b_{d_2}\}$.
	Note that the state $\a_1\ox \b_1$ is compatible with a bipartite entangled state $\tau=2\ve^2 (\ket{00}+\ket{11})(\bra{00}+\bra{11})$. Hence $\s_1$ is compatible with a state
	\begin{eqnarray}
	\s_2=\tau+ \a_1\ox \b_2+ \a_2\ox (\b_1+\b_2),
	\end{eqnarray}
	and $\s_1\neq \s_2$.
	That is, $\r$ is compatible with two different states $\s_1$ and $\s_2$. This shows that $\r$ is not 1-UDA.
	This completes the proof.
\end{proof}
It has been shown that if one of $\r_{A_1}$ and $\r_{A_2}$ is pure, then $\r$ is 1-UDA. 
Proposition \ref{pro:bipartite1-UDA} further shows that this is the only possible condition that $\r$ is 1-UDA. Therefore, the UDA property requires that the rank of the density operator is relatively low.

\subsection{On tripartite mixed UDA states}
\label{sec:tripartite_state}
We characterize a three-qubit mixed 2-UDA state $\r_{A_1}\ox \r_{A_2A_3}$ by deriving the necessary and sufficient condition it satisfies. By this fact, one can determine whether such a state is 2-UDA directly. It is helpful for the analysis of the additivity of mixed UDA states.

To begin with,  we consider a three-qubit state $\r$ in the form of $\r_{A_1}\ox \r_{A_2A_3}$. Obviously, if 
one of the six 1- and 2-marginals of $\r$ is pure, then $\r$ is 2-UDA.
In fact,  we assume that $\r_{A_1}$ or $\r_{A_2A_3}$ is pure. Then $\r=\proj{\eta}_{A_1} \ox\r_{A_2A_3}$  (resp.  $\r=\r_{A_1} \ox \proj{\ps}_{A_2A_3}$). 
By Lemma \ref{le:pure sep}, we obtain that a state $\s$ that is 2-compatible with $\r$ satisfies that 
$\s=\proj{\eta}_{A_1} \ox \s_{A_2A_3}$ 
 (resp. $\s=\s_{A_1} \ox \proj{\ps}_{A_2A_3}$).
By $\r_{A_2A_3}=\s_{A_2A_3}$ (resp. $\r_{A_1A_2}=\s_{A_1A_2}$), we have  $\r=\s$ and $\r$ is 2-UDA. Conversely, a 2-UDA state  $\r_{A_1}\ox \r_{A_2A_3}$ does not necessarily satisfy that one of the six 1- and 2-marginals of $\r$ is pure. Here is an example. 
\begin{example}
\label{exp:xi^(x)}
We consider the state
	$\xi^{(x)}_{A_1A_2A_3}=(\frac{1}{2}I_2)_{A_1}\ox \xi_{A_2A_3}^{(x)}$ with
	$
\xi^{(x)}_{A_2A_3}=\frac{1}{3}\oplus \bma 
	\frac{1}{3} & x \\
	x^* & \frac{1}{3} \ema \oplus 0
	$.
	By choosing $x=1/3$, one can verify that  $\xi^{(1/3)}$ is 2-UDA. Besides, we have  $\rank(\xi^{(1/3)}_{A_1A_j})=4$ for $j =2,3$ and
	$\rank(\xi^{(1/3)}_{A_2A_3})=\rank(\xi^{(1/3)}_{A_k})=2$ for $k=1,2,3$, i.e. neither of the 1- and 2-marginals of $\xi^{(1/3)}$ is pure. 
	Further we have $\xi^{(x)}$ is 2-UDA iff $\abs{x}=\frac{1}{3}$.
\end{example}
Inspired by this example, we consider the necessary and sufficient condition with which the state $\r=\r_{A_1}\ox \r_{A_2A_3}$ is 2-UDA. 
Our analysis is presented in a stepwise manner according to the rank of $\r_{A_2A_3}$. Several cases for rank-two $\r_{A_2A_3}$ and rank-three $\r_{A_2A_3}$ LU equivalent to a state with one zero diagonal element are presented in Proposition \ref{pro:D+0iff|D|=0}. Subsequently, other rank-three  $\r_{A_2A_3}$ and rank-four $\r_{A_2A_3}$ are analyzed in Proposition \ref{pro:rankA23=3or 4}.    To highlight the main results in Theorem \ref{th:rhoA1 ox A23 iff}, these intermediate results and proofs are provided in Appendix \ref{app:3-qubit}.
By summarizing the preceding facts, we obtain the complete characterization of a 2-UDA product  state $\r_{A_1}\ox \r_{A_2A_3}$ as follows.
\begin{theorem}
	\label{th:rhoA1 ox A23 iff}
	Let $\r=\r_{A_1}\ox \r_{A_2A_3}$ be a three-qubit state. Then $\r$ is 2-UDA if and only if $\r$ satisfies one of the following conditions,
	
	(i) at least one of the 1-marginals  of $\r$  is pure;
	
	 (ii) $\r_{A_2A_3}$ is pure;	
	
	(iii) $\rank(\r_{A_1})=\rank(\r_{A_2A_3})=2$ and $\r_{A_2A_3}$ is LU equivalent to $D\oplus 0$ in  \eqref{eq:rhoA23=D+0} with $q_1q_2q_3>0$.
\end{theorem}

\begin{proof}
First, we claim that if $\rank(\r_{A_1})=1$ or $\rank(\r_{A_2A_3})=1$, then $\r$ is 2-UDA. The proof is trivial by Lemma \ref{le:pure sep}. 
Second, if  $\rank(\r_{A_1})=2$ and $\rank(\r_{A_2A_3})=2$, then $\r$ is LU-equivalent to $\r_{A_1}\ox (D\oplus 0)_{A_2A_3}$ in  \eqref{eq:rhoA23=D+0}. If $q_1q_2q_3>0$ in  \eqref{eq:rhoA23=D+0}  or $q_2=0$ or $q_3=0$ in $D$, then $\r$ is 2-UDA; If $q_1=0$, then $\r$ is not 2-UDA.
The first claim is derived by Proposition \ref{pro:D+0iff|D|=0}. Next,
 If  $q_2=0$ or $q_3=0$ in $D$, then $\r$ is 2-UDA by Proposition \ref{pro:D+0iff|D|=0} (i).
 if $q_k>0$, then $\rank(\r_{A_2A_3})=2$ implies that $\abs{D}=0$, and $\r$ is 2-UDA by Proposition \ref{pro:D+0iff|D|=0} (ii). If $q_1=0$ and $\rank(\r_{A_2A_3})=2$, then $\abs{x}<\sqrt{q_2q_3}$, and $\r$ is not 2-UDA by Proposition \ref{pro:D+0iff|D|=0} (iii).
Finally, if  $\rank(\r_{A_1})=2$ and $\rank(\r_{A_2A_3})=3 \text{ or } 4$ , then $\r$ is not 2-UDA. In fact, if $\r_{A_2A_3}$ is LU equivalent to a state that has a zero diagonal element, all the remaining diagonal elements are strictly larger than zero, and  $\r$ is not 2-UDA by Proposition \ref{pro:D+0iff|D|=0} (ii). Otherwise,  $\r$ is not 2-UDA by Proposition \ref{pro:rankA23=3or 4} (i). The result for full-rank $\r_{A_2 A_3}$ is directly obtained by Proposition \ref{pro:rankA23=3or 4} (ii).
\end{proof}

Theorem \ref{th:rhoA1 ox A23 iff}  gives an available criterion for when two-body tomography is sufficient for this natural three-qubit family. For the mixed product family $\rho_{A_1}\otimes \rho_{A_2A_3}$, the 2-UDA property is a boundary phenomenon rather than a generic one: although the 2-marginals determine $\rho_{A_1}$ and $\rho_{A_2A_3}$, they do not  rule out another global state with the same marginals, because off-diagonal blocks of the form $\lvert i\rangle\langle j\rvert_{A_1}\otimes X_{ij}$ may disappear under all two-body partial traces.

For the diagonal state $\r_{A_2A_3}$, the result in Theorem \ref{th:rhoA1 ox A23 iff} can be extended to arbitrary finite dimension. 
\begin{proposition}
	\label{pro:rho2-udaDiagonald-dim}
	Suppose $\r=\r_{A_1}\ox \r_{A_2A_3}$ is a tripartite state and $\r_{A_2A_3}$ is diagonal. Then $\r$ is 2-UDA if and only if  one of the six 1- and 2-marginals of $\r$ is pure.
\end{proposition}

The proof of Proposition \ref{pro:rho2-udaDiagonald-dim} is shown in Appendix \ref{app:3-qubit}. For arbitrary states, the family of 2-UDA states exhibits a much more complicated structure, so the purity of the six 1- and 2-marginals can only serve as a sufficient condition and not a full characterization in Theorem \ref{th:rhoA1 ox A23 iff}. By contrast, for states with diagonal $\rho_{A_2A_3}$, the additional structural simplification makes the 2-UDA property much more tractable. In this case, the same condition becomes necessary and sufficient in Proposition \ref{pro:rho2-udaDiagonald-dim}, showing that the diagonal structure captures the essential features relevant to 2-UDA.

For arbitrary three-qubit or tripartite states, the complete characterization of 2-UDA states is much more complicated. 
As indicated by the pattern observed in Theorem \ref{th:rhoA1 ox A23 iff}, the more mixed a state is, the less likely it is to be 2-UDA. There may therefore exist a critical rank (for instance, $\rank(\rho_{A_1})=\rank(\r_{A_2A_3})=2$ in Theorem \ref{th:rhoA1 ox A23 iff}) at which a transition occurs: states at this rank may or may not be 2-UDA, whereas all states of smaller rank are 2-UDA and all states of larger rank are not. The following result shows that if  the rank of $\rho$ is six with additional conditions, or greater than seven, then $\rho$ is necessarily not 2-UDA. 
\begin{proposition}
\label{pro:3qubit_general}
A three-qubit state $\r_{A_1A_2A_3}$ with $\rank(\r)=6$ and $\rank(\r_{A_2A_3})= 3$ is not 2-UDA.  Further, $\r_{A_1A_2A_3}$ with rank at least seven is not 2-UDA. 
\end{proposition}

The proof of Proposition \ref{pro:3qubit_general} is shown in Appendix \ref{app:3-qubit}. 
We will extend the second result to a $n$-qubit state $\r$ supported on $\bbC^{d_1}\ox \bbC^{d_2}\ox ... \ox \bbC^{d_n}$ in Proposition \ref{pro:rankn-1}.  
Besides, we conjecture that the critical rank of a three-qubit state ranges from four to six. 
The complete characterization of a general tripartite mixed state remains to be further explored.

\subsection{On four-qubit mixed UDA states}
\label{sec:four-qubit}
By virtue of the essential characterization of the three-qubit 2-UDA state, we present the necessary and sufficient condition by which a mixed state $\r=\r_{A_1A_2}\ox \r_{A_3A_4}$ is 2-UDA. It is shown that it is possible for the state to be 2-UDA only when the rank of $\r$ is less than six. We further deal with the "low rank" case and show the necessary and sufficient condition in Theorem \ref{th:RA12oxRA34}.

As the first step, we assume that $\r_{A_1A_2}$ is a product state, say $\r=\r_{A_1}\ox \r_{A_2}\ox\r_{A_3A_4}$. By detailed discussions, we obtain the complete characterization of the 2-UDA state $\r$   in Proposition \ref{pro:4-qubit2UDA}. That is, if at least one of $\r_{A_1}$ or $\r_{A_2}$ is pure, then $\r$ is 2-UDA iff the remaining three-qubit state $\r_{A_k}\ox \r_{A_3A_4}$ is 2-UDA given in Theorem \ref{th:rhoA1 ox A23 iff}. Otherwise, $\r$ is 2-UDA iff one of $\r_{A_3A_4}$ is pure.
Second, the 2-UDA property of states $\r=\r_{A_1A_2}\ox \r_{A_3A_4}$ with $\rank(\r_{A_1A_2})=\rank(\r_{A_3A_4})=2$ is fully analyzed in Proposition \ref{pro:rank4_rA12OxrA34}, depending on the form of $\r_{A_1A_2}$. 
Finally, we analyze the 2-UDA property of  high-rank product states in Proposition \ref{pro:4qubit-highrank}, i.e. if one of $\rank(\r_{A_1A_2})$ and $\rank(\r_{A_3A_4})$ has rank not less than three, and the other is not pure, then $\r=\r_{A_1A_2}\ox \r_{A_3A_4}$ is not 2-UDA.

By now, we have completed the analysis of the 2-UDA property
of any four-qubit state $\r=\r_{A_1A_2}\ox \r_{A_3A_4}$. We conclude the results of Propositions \ref{pro:4-qubit2UDA}-\ref{pro:4qubit-highrank}, and obtain the necessary and sufficient condition by which  $\r=\r_{A_1A_2}\ox \r_{A_3A_4}$ is 2-UDA as follows.
\begin{theorem}
\label{th:RA12oxRA34}
    Let $\r=\r_{A_1A_2}\ox \r_{A_3A_4}$ be a four-qubit state. Then $\r$ is 2-UDA if and only if one of the following conditions holds,

(i) $\r_{A_1A_2}$ or $\r_{A_3A_4}$ is pure;

(ii) at least one of $\{\r_{A_1},\r_{A_2}\}$ and $\{\r_{A_3},\r_{A_4}\}$ are pure, respectively;

(iii) only one of the 1-marginals $\r_{A_j}$ is pure, and $\r=\r_{A_j}\ox \r_{A_n}\ox \r_{A_{m_1}A_{m_2}}$ satisfies that $\rank(\r_{A_{m_1}A_{m_2}})=2$ and $\r_{A_{m_1}A_{m_2}}$ is LU equivalent with $D\oplus 0$ in  \eqref{eq:rhoA23=D+0} with $q_1q_2q_3>0$.

(iv) $\r_{A_1A_2}$
   and $\r_{A_3A_4}$  are rank-two, and their product is LU equivalent to
\begin{eqnarray}
\label{eq:D+0oxH+0}
&&(H\oplus0)_{A_1A_2} \ox (D\oplus0)_{A_3A_4}
\\
&=&\bma p_1 &b & c & 0\\
b^* &p_2&a &0 \\
c^*& a^* &p_3&0\\
0& 0& 0& 0\ema\ox 
\bma q_1 &y & z & 0\\
y^* &q_2&x &0 \\
z^*& x^* &q_3&0\\
0& 0& 0& 0\ema,
\nonumber
\end{eqnarray}
where $p_1p_2p_3>0$ and $q_1q_2q_3>0$. 
\end{theorem}

Theorem \ref{th:RA12oxRA34} is a structural classification  for the 2-UDA property within the class of four-qubit product states $\rho_{A_1 A_2} \otimes \rho_{A_3 A_4}$. It converts an a priori global uniqueness problem into a finite set of local and available conditions. The appearance of LU-equivalence classes in the rank-two case imply that the boundary between UDA and non-UDA states is determined by an intrinsic physical structure rather than by a basis-dependent description. For quantum tomography or state certification, one may first reconstruct the relevant 1- and 2-marginals, evaluate their purities and ranks, and then directly decide whether the global state is already uniquely fixed by 2-marginals or whether full four-qubit tomography is still necessary.

\subsection{On $n$-qubit mixed UDA states for $n\geq 5$}
\label{sec:n-qubit}
We show the necessary and sufficient conditions under which  some $n$-qubit states are $k$-UDA. We introduce a systematic method for the determination of non-$k$-UDA states in FIG. \ref{fig:k-UDA}, which derives the construction of multipartite states that is not $k$-UDA.

To begin with, we show the complete characterization of a 2-UDA five-qubit state $\r=\r_{A_1}\ox \r_{A_2A_3}\ox \r_{A_4A_5}$.  We show that $\r$ is not 2-UDA if  $\r_{A_1}$ is mixed and $\r_{A_2A_3}$ and $\r_{A_4A_5}$ satisfy the condition in Theorem \ref{th:RA12oxRA34} (iv), which is shown in Lemma \ref{le:rA1oxrA2345}.
Combining with Theorems  \ref{th:rhoA1 ox A23 iff} and \ref{th:RA12oxRA34}, Proposition \ref{pro:n-qubit} and Lemma \ref{le:rA1oxrA2345}, we have the complete characterization of a five-qubit product state. 
\begin{theorem}
\label{th:5-qubit}
The five-qubit state $\r=\r_{A_1}\ox \r_{A_2A_3}\ox \r_{A_4A_5}$ is 2-UDA if and only if one of the following conditions holds, 

(i) $\r_{A_2A_3}$ or $\r_{A_4A_5}$ is pure, i.e. $\r=\r_{A_1}\ox \proj{\ps}_{A_{n_1}A_{n_2}}\ox \r_{A_{m_1}A_{m_2}}$, and the remaining state $\r_{A_1}\ox \r_{A_{m_1}A_{m_2}}$ satisfies the conditions in Theorem \ref{th:rhoA1 ox A23 iff}.

(ii) $\r_{A_1}$ is pure, and at least one of $\{\r_{A_2},\r_{A_3}\}$ and $\{\r_{A_4},\r_{A_5}\}$ are pure, respectively.

(iii) Up to LU equivalence, $\r=\proj{\ps}_{A_1}\ox \proj{\phi}_{A_{n_1}}\ox \r_{A_{n_2}}\ox (D\oplus 0)_{A_{m_1}A_{m_2}}$, where $\{n_1,n_2\}= \{2,3\}$ or $\{4,5\}$, $\r_{A_{n_2}}$ is rank-two, and $D$ is given in \eqref{eq:rhoA23=D+0}. 

(iv) Up to LU equivalence, $\r=\proj{\ps}_{A_1}\ox (H\oplus 0)_{A_{2}A_{3}}\ox 
(D\oplus 0)_{A_{4}A_{5}}$, where $H$ and $D$ are given in \eqref{eq:D+0oxH+0}. 
\end{theorem}
\begin{proof}
Obviously, $\r$ is 2-UDA only if $\r_{A_2A_3}\ox \r_{A_4A_5}$ is 2-UDA by Theorem \ref{th:RA12oxRA34}. This derives cases (i)-(iv) as follows. 
(i) $\r_{A_2A_3}$ or $\r_{A_4A_5}$ is pure. Then $\r$ is 2-UDA iff the remaining part is 2-UDA by Lemma \ref{le:pure sep}. 
(ii) At least  one of $\{\r_{A_2},\r_{A_3}\}$ and $\{\r_{A_4},\r_{A_5}\}$ are pure, respectively. Then $\r=\r_{A_1}\ox \proj{\ps}_{A_{n_1}}\ox \r_{A_{n_2}}\ox \proj{\phi}_{A_{m_1}}\ox \r_{A_{m_2}}$ with mixed $\r_{A_{n_2}}$ and $\r_{A_{m_2}}$. Using Theorem \ref{th:RA12oxRA34}, $\r$ is 2-UDA iff $\r_{A_1}\ox \r_{A_{n_2}}\ox \r_{A_{m_2}}$ is 2-UDA, i.e. $\r_{A_1}$ is pure.  
(iii) Up to LU equivalence, $\r_{A_2...A_5}=\proj{\phi}_{A_{n_1}}\ox \r_{A_{n_2}}\ox (D\oplus 0)_{A_{m_1}A_{m_2}}$.
$\r$ is 2-UDA iff $\r_{A_1}$ is pure by Theorem \ref{th:RA12oxRA34}. Otherwise, $\r_{A_1}\ox \r_{A_{n_2}}\ox (D\oplus 0)_{A_{m_1}A_{m_2}}$ is not 2-UDA. 
(iv) Up to LU equivalence, $\r_{A_2...A_5}= (D\oplus 0)_{A_{2}A_{3}}\ox 
(H\oplus 0)_{A_{4}A_{5}}$. By Lemma \ref{le:rA1oxrA2345}, $\r$ is 2-UDA iff $\r_{A_1}$ is pure. 
\end{proof}

Theorem \ref{th:5-qubit} reveals a clear structural picture of five-qubit product-structured mixed 2-UDA states. Uniqueness is supported by only two mechanisms. First, sufficiently many pure 1-marginals, which enforce successive factorization and reduce the five-partite problem recursively to lower-partite ones. Second, a very small class of  rank-two two-body blocks. From this perspective, the five-qubit result  shows that higher-partite 2-UDA relies on lower-partite 2-UDA results and is easily destroyed by additional mixed local degrees of freedom. This is contrast with the pure-state setting, where the UDA property is often generic, whereas in the mixed-state regime 2-UDA survives only in highly constrained low-rank configurations.

We present the complete characterization of some $n$-qubit mixed $k$-UDA states. First, we consider the fully product $k$-UDA state $\ox_{i=1}^n\r_{A_i}$. On the basis of that, we consider two general cases in (ii) and (iii) for 2-UDA states.
\begin{proposition}
\label{pro:n-qubit}
(i)  The  $n$-partite state $\r=\ox_{i=1}^n\r_{A_i}$ is $k$-UDA if and only if at least $n-k$ of the 1-marginals $\r_{A_i}$ is pure.  

(ii) The  $n$-qubit state $\r=\ox_{i=1}^{n-2}\r_{A_i}\ox \r_{A_{n-1}A_n}$ is $2$-UDA if and only if one of the following conditions holds, 
(a) $\r_{A_{n-1}A_n}$ is pure, and $\r=\ox_{i=1}^{n-2}\r_{A_i}$ satisfies (i);
(b) $\r_{A_{n-1}}$ or $\r_{A_{n}}$ is pure, then $\r=\ox_{i=1}^n\r_{A_i}$, which satisfies (i); 
(c) $\ox_{i=1}^{n-2}\r_{A_i}$ is pure, and $\r_{A_{n-1}A_n}$ is arbitrary;
(d) at least $n-3$ of $\r_{A_i}$ is pure, $\r_{A_{n-1}A_n}$ is rank-two and  LU equivalent with $D\oplus 0$ in  \eqref{eq:rhoA23=D+0} with $q_1q_2q_3>0$.

(iii) The  $n$-qubit state $\r=\ox_{i=1}^{n-4}\r_{A_i}\ox \r_{A_{n-3}A_{n-2}}\ox \r_{A_{n-1}A_n}$ is $2$-UDA if and only if one of the following conditions holds, 
(a) $\r_{A_{n-3}A_{n-2}}$ or
$\r_{A_{n-1}A_{n}}$ is pure, and $\ox_{i=1}^{n-4}\r_{A_i}\ox \r_{A_{j_1}A_{j_2}}$ satisfies (ii); (b) at least one of $\r_{A_{j}}$ is pure, for $j=n-3,...,n$, and $\r=\ox_{i=1}^{n-4}\r_{A_i}\ox \r_{A_{j}}\ox \r_{A_{l}}\ox \r_{A_{m_1}A_{m_2}}$ satisfies (ii);
(c) $\r_{A_{n-3}A_{n-2}}$ and $\r_{A_{n-1}A_{n}}$ are rank-two, 
$\r_{A_{n-3}A_{n-2}}\ox \r_{A_{n-1}A_n}$ is LU equivalent to  $(H\oplus0) \ox (D\oplus0)$ in \eqref{eq:D+0oxH+0}, and $\ox_{i=1}^{n-4}\r_{A_i}$ is pure. 
\end{proposition}

The proof of Proposition \ref{pro:n-qubit} is given in Appendix \ref{app:n-qubit}. 
Proposition \ref{pro:n-qubit} shows that the dependence on $k$ is structurally transparent. For the fully product $k$-UDA state in (i), it allows up to $k$ local qubits to remain mixed. Consequently, the admissible class of $k$-UDA states becomes larger as $k$ increases; at the same time, the maximal compatible rank also grows. Moreover, the family is nested in $k$, i.e., any state that is $k$-UDA is automatically $\ell$-UDA for every $\ell\ge k$, since the $\ell$-marginals determine the $k$-marginals by partial trace. By contrast, the $2$-UDA classifications in Proposition \ref{pro:n-qubit} (ii) and (iii) show that once nontrivial two-body correlated blocks are present, the additional freedom is much more limited. Further, we introduce some $n$-qubit 2-UDA states. An $n$-qubit state $\r$ is a symmetric state if $\r=\text{Swap}_{(i,j)}\r$ for all $1\leq i,j\leq n$. It is shown that if all 2-marginals of an $n$-qubit state $\r$ are symmetric, then $\r$ is symmetric \cite{shi2025entangle}. Actually, we only need $n-1$ 2-marginals corresponding to a connected graph, say $\rho_{12}, \rho_{23},..., \rho_{(n-1)n}$. Hence, the multiqubit Dicke state is 2-UDA.

We turn to consider the UDA property of general $n$-partite states $\r_{A_1A_2...A_n}$ supported on $\bbC^{d_1}\ox \bbC^{d_2} \ox ...\ox \bbC^{d_n}$. It is shown that some mixed states with high rank is definitely not $k$-UDA even with large $k$. 
\begin{proposition}
\label{pro:rankn-1}
(i) Every  $n$-partite mixed state $\r$ supported on $\bbC^{d_1}\ox \bbC^{d_2}\ox...\ox \bbC^{d_n}$ with $\rank(\r)\geq d_1d_2...d_n-1$ is not $k$-UDA, for $1\leq k\leq n-1$.  Besides, the full-rank $\r$ can be compatible with another full rank state.  

(ii) The $n$-qubit state $\r_{A_1A_2...A_n}$ with $\rank(\r)=d_1(d_2d_3...d_n-1)$ and $\rank(\r_{A_2...A_n})= d_2d_3...d_n-1$, then the state is not $k$-UDA, for $1\leq k\leq n-2$. 
\end{proposition}

The proof of Proposition \ref{pro:rankn-1} is given in Appendix \ref{app:n-qubit}.  It shows that high rank itself already constitutes a fundamental obstruction to be $k$-UDA. In particular, Proposition \ref{pro:rankn-1} (i) implies that local marginals become highly non-informative in the near-full-rank regime.  Proposition \ref{pro:rankn-1} (ii) shows a sharper criterion in the $n$-qubit case, where the failure of $k$-UDA already appears under a more structured rank condition. These facts delineate the boundary of applicability of the $k$-UDA framework and suggest that uniqueness is more likely to hold only for low-rank or specially structured states.

With respect to any unitarily invariant induced measure $\mu_{N,K}$ with $K\ge N$
(in particular the Hilbert--Schmidt measure),
the set of rank-deficient density matrices has measure zero \cite{Zyczkowski2001induced}.
 This implies the following fact.
\begin{corollary}
\label{cor:almost}
Almost all $n$-partite mixed states are not $k$-UDA. 
\end{corollary}
\begin{proof}
Let
$\cH=\bigotimes_{i=1}^n \mathbb C^{d_i}$,
 $D:=\prod_{i=1}^n d_i$
and
$\cD(\cH)$
be the set of quantum states on $\mathcal H$. For every $1\le k\le n-1$, we show that the set of $k$-UDA states has measure zero in $\mathcal D(\mathcal H)$ with respect to the $(D^2-1)$-dimensional Lebesgue measure induced by the Hilbert--Schmidt metric.
Consider the affine hyperplane
$\mathcal A:=\{X\in \mathrm{Herm}(\mathcal H): \tr(X)=1\}
\cong \mathbb R^{D^2-1}$,
equipped with the Lebesgue measure induced by the Hilbert-Schmidt inner product.
Let 
$\mathcal U_k
:=
\{\rho\in\mathcal D(\mathcal H): \rho \text{ is } k\text{-UDA}\}$.
By Proposition \ref{pro:rankn-1} (i), if $\rho$ is $k$-UDA, then 
$\rank(\rho)\le D-2$.
Hence,
$\mathcal U_k
\subseteq
\{\rho\in\mathcal D(\mathcal H): \rank(\rho)\le D-2\}
\subseteq
\{\rho\in\mathcal D(\mathcal H): \det(\rho)=0\}$.
 It follows that
$\mu(\mathcal U_k)
\le
\mu(Z\cap\mathcal D(\mathcal H))
=0$,
where $Z:=\{\rho\in\mathcal A:\det(\rho)=0\}$, and $\mu$ denotes the induced Lebesgue measure on $\mathcal D(\mathcal H)$.
Therefore, the set of $k$-UDA states has measure zero. 
\end{proof}

Compared with the pure-state UDA theory, the mixed-state picture obtained here is different rather than merely a technical extension. In the pure-state setting, low-order marginals can determine the global state generically, for example, almost all three-qubit pure states are 2-UDA, with GHZ-type states forming the exceptional non-UDA family \cite{den2002almost,walck2008only}. By contrast, Theorems \ref{th:rhoA1 ox A23 iff}, \ref{th:RA12oxRA34}, and \ref{th:5-qubit} show that, within the mixed-state product families considered here, 2-UDA survives only in highly constrained low-rank regimes, typically supported by pure one-body marginals or by rigid rank-two canonical blocks. Proposition \ref{pro:rankn-1} and Corollary \ref{cor:almost} further show that sufficiently high rank rules out
$k$-UDA and that almost all mixed states are not 
$k$-UDA. Hence the mixed-state problem is governed not only by entanglement structure, but also by rank, positivity, and the convex geometry of the state space. 

\subsection{Systematic determination of mixed  UDA states and experimentally relevant families}

\begin{figure}[h]
    \centering
\includegraphics[width=1\linewidth]{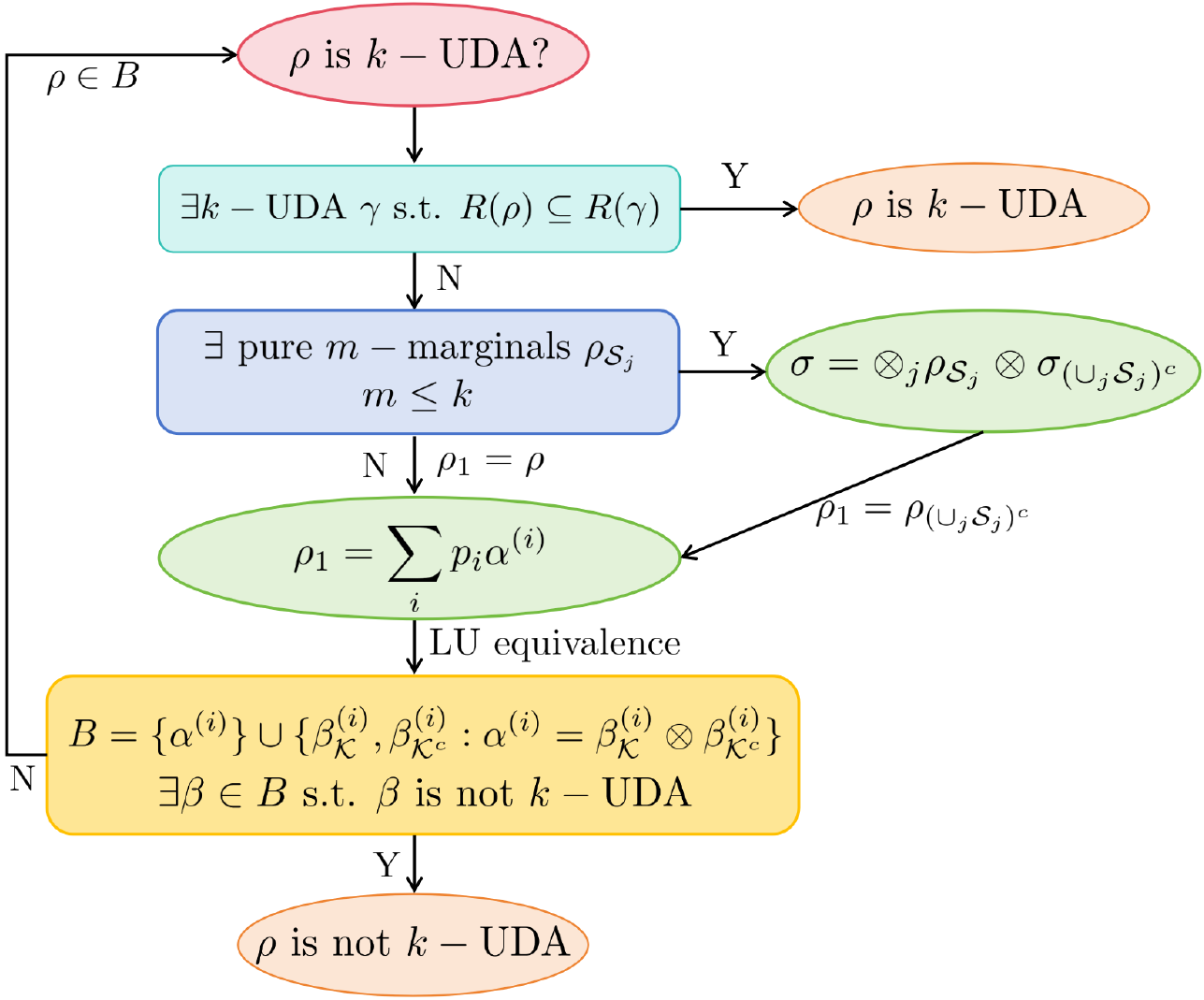}
    \caption{Procedure for the determination of $k$-UDA states. Here $\s$ is any state $k$-compatible with $\r$. The first to third step are shown in the cyan, blue and yellow boxes, respectively. }
    \label{fig:k-UDA}
\end{figure}

The determination of $n$-qubit states that are $k$-UDA is an essential task in quantum state tomography. The full characterization of bipartite to $n$-qubit states we presented in this paper provides a systematic determination of $k$-UDA states, as shown in FIG \ref{fig:k-UDA}. This facilitates narrowing the scope of $k$-UDA states, thereby streamlining their verification and construction.  
Our goal is to determine  more $k$-UDA states (not necessarily product states) by known results, and transform a state on more systems and with higher rank into a set of its "components", i.e., the marginals of the global state or the elements in its range. Obviously, these states on fewer systems or with lower rank are easier to deal with, as our complete characterization of some multipartite results. The procedure consists of  three steps as follows.
\begin{itemize}
\item Step 1 (Determination by range inclusion) 
Let $\r$ be a global state to be determined.  By Lemma \ref{le:lr+(1-l)s_not2uda} (ii), if $\r$ is an element of  the face of $\cR(\g)$ and $\g$ is $k$-UDA, then $\rho$ is $k$-UDA. Otherwise, $\r$ is turned to the non-$k$-UDA verification. 
 
 \item  Step 2 (Reduction of systems):  All  $m$-marginals of $\r$ are verified by their rank, for $m\leq k$. If the $m$-marginals $\r_{\cS_j}$ are pure, then any state $k$-compatible with $\r$ is given by $\s=\ox_j \r_{\cS_j}\ox \s_{(\cup_j \cS_j)^c}$. Then $\r$ is $k$-UDA iff the remaining part of $\s$, i.e. $\s_{(\cup_j \cS_j)^c}$ is fully determined. That is, the state $\r_1:=\r_{(\cup_j \cS_j)^c}$ on fewer systems is $k$-UDA.
Otherwise, if none of the $m$-partite marginals of $\r$ is pure, we choose $\r_1=\r$. 

\item Step 3 (Reduction of rank): $\r_1$ is decomposed to the convex combination of some states $\a^{(i)}$, i.e. $\r=\sum_i p_i \a^{(i)}$. The rank of $\a^{(i)}$ will be less than that of $\r_i$ for most cases. The simplification of $\a^{(i)}$ is applied by LU equivalence. The set $B$ of states to be verified is constructed as follows.
If $\a^{(i)}$ is a product state, i.e. $\a^{(i)}=\b^{(i)}_\cK\ox \b^{(i)}_{\cK^c}$, then 
$B=\{\a^{(i)}\}\cup \{\b^{(i)}_\cK, \b^{(i)}_{\cK^c}\}$; otherwise, $B=\{\a^{(i)}\}$.  
Using Lemma \ref{le:lr+(1-l)s_not2uda} (i) and (ii), if there is an element of $B$ that is not $k$-UDA by the necessary and sufficient conditions in Propositions \ref{pro:bipartite1-UDA} and \ref{pro:n-qubit}, and Theorems \ref{th:rhoA1 ox A23 iff}, \ref{th:RA12oxRA34} and \ref{th:5-qubit}, we conclude that $\r$ is not $k$-UDA; otherwise, we choose each element $B$ as $\r$, and repeat STEP 1 to find possible states that are not $k$-UDA. 
\end{itemize}
From this procedure, one can see that this method can  partially conclude or exclude that a state is $k$-UDA.
Even if no non-$k$-UDA component is detected through this procedure, the global state $\r$ may still fail to be $k$-UDA.
What can be stated with certainty is that this approach significantly reduces the complexity of assessing whether a multipartite mixed state is $k$-UDA by decomposing a high-rank many-body quantum state into a sequence of lower-rank, smaller subsystem components. Further, this procedure provides a systematic construction of multipartite states that are not $k$-UDA. This method implies that a state on fewer systems is vital for determining the $k$-UDA property, which is exactly what we have shown in this work.    

To illustrate how this procedure works in practice, we now examine three families of mixed states, exhibiting both $2$-UDA and non-$2$-UDA behavior. 
\begin{example}

We denote 
$|\Psi^+\rangle := \frac{|01\rangle+|10\rangle}{\sqrt2},
\;
|\Phi^+\rangle := \frac{|00\rangle+|11\rangle}{\sqrt2}$ and $|\Phi_L^\pm\rangle
:=
(|0\rangle|00\rangle \pm |1\rangle|\Psi^+\rangle)/\sqrt2
$. 
Recall that for $0\leq x\leq \frac{1}{3}$, 
$\xi^{(x)}_{A_iA_jA_k}:=\left(\frac{I_2}{2}\right)_{A_i} \otimes \xi^{(x)}_{A_jA_k}$  in Example \ref{exp:xi^(x)} 
 is 2-UDA if and only if $x=\frac13 $.
We present three families of mixed states determined by the procedure.
\begin{itemize}
\item The  mixed state family
$\a_{p}
=
p|\Phi_L^+\rangle\!\langle\Phi_L^+|
+
(1-p)|\Phi_L^-\rangle\!\langle\Phi_L^-|$  represents a logical Bell state subjected to phase-flip noise.
Then $\a_p$ with $0<p<1$ is 2-UDA. 
In fact, we have 
$\cR(\a_{p})
=
\operatorname{span}\Bigl\{|0\rangle|00\rangle,\,
|1\rangle|\Psi^+\rangle\Bigr\}
\subseteq 
\cR(\xi^{(1/3)})$.
Since $\xi^{(1/3)}$ is 2-UDA, $\a_{p}$ is $2$-UDA by Step 1.

\item The  five-qubit state
$\Omega_{p,x}
=
|\Phi^+\rangle\!\langle\Phi^+|_{A_1A_2}\otimes \big(p|\Phi_{L}^+\rangle\!\langle\Phi_{L}^+|
+(1-p)\xi^{(x)}\big)_{A_3A_4A_5}
$ for $0<p<1$ describes a composite entanglement resource consisting of an ideal Bell pair on $A_1A_2$ and a noisy three-qubit logical block on $A_3A_4A_5$.
Then $\Omega_{p,x}$ is 2-UDA if and only if $x=\frac{1}{3}$. In fact, the 2-marginal of $\Omega_{p,x}$ on $A_1A_2$ is pure. By Step 2, any state $\sigma$ that is $2$-compatible with $\Omega_{p,x}$ must be of the form
$\sigma
=
|\Phi^+\rangle\!\langle\Phi^+|_{A_1A_2}\otimes \sigma_{A_3A_4A_5}$.
Now we analyze the remaining three-qubit state. If $x=1/3$, then
$\cR\Big((\Omega_{p,\frac{1}{3}})_{A_3A_4A_5}\Big)\subseteq \cR(\xi^{(1/3)})$.
Then $(\Omega_{p,\frac{1}{3}})_{A_3A_4A_5}$ is $2$-UDA by Step 1, and hence $\Omega_{p,1/3}$ is $2$-UDA.
If $0\le x<1/3$, then $\xi_{A_3A_4A_5}^{(x)}$ is not $2$-UDA. Since this non-$2$-UDA state appears in the convex decomposition of $(\Omega_{p,x})_{A_3A_4A_5}$ with nonzero weight $1-p$, Step 3  implies that $(\Omega_{p,x})_{A_3A_4A_5}$ is not $2$-UDA. Hence $\Omega_{p,x}$ for $0<p<1$ and $0\le x <\frac{1}{3}$ is not $2$-UDA either.
\item 
Let $P_n$ be the path graph on $n\ge 4$ vertices, and $\ket{LC_n}
=
\Bigl(\prod_{i=1}^{n-1} CZ_{i,i+1}\Bigr)\ket{+}^{\otimes n}$ be the corresponding linear cluster state. For each $x=(x_1,\dots,x_n)\in\{0,1\}^n$, we define
$
Z^x:=Z_1^{x_1}\cdots Z_n^{x_n},
\;
\ket{LC_x}:=Z^x\ket{LC_n}.
$
Then the locally dephased graph-diagonal family
$$
\rho_{\mathbf q}
=
\sum_{x\in\{0,1\}^n}
\Bigl(\prod_{i=1}^n q_i^{x_i}(1-q_i)^{1-x_i}\Bigr)
\proj{LC_x},
$$ for $0\le q_i\le 1$ is not 2-UDA.
In fact, Step 1 and 2 give no useful information, while
the canonical decomposition above is exactly the input required by Step 3. Since all pure components $\proj{LC_x}$ are locally unitary equivalent, it suffices to test
$\alpha^{(0)}:=\ket{LC_n}\!\bra{LC_n}
=
\frac{1}{2^n}\sum_{g\in\cS} g$, where $\cS$ is the stabilizer group of $\ket{LC_n}$  given by $\mathcal S=\langle K_1,\dots,K_n\rangle$ with  generators  $K_1=X_1Z_2$,
$K_i=Z_{i-1}X_iZ_{i+1}$ for $2\le i\le n-1$ and
$K_n=Z_{n-1}X_n$.
Let
$\s=\frac{1}{2^{n-2}}
\bigl(\frac{I+K_1}{2}\bigr)\bigl(\frac{I+K_n}{2}\bigr)$ satisfying $\s\ge 0$ and $\operatorname{Tr}(\s)=1$.
One can verify that
$\s_{ij}=\alpha^{(0)}_{ij}$,
$\forall\,1\le i<j\le n$.
Since $\s\neq \alpha^{(0)}$, the pure cluster state $\alpha^{(0)}$ is not $2$-UDA.
By local unitary equivalence, each $\alpha^{(x)}$ is also not $2$-UDA. 
Hence, $\rho_{\mathbf q}$ is not $2$-UDA.
\end{itemize}

\end{example}
The examples show that the determination procedure is not restricted to the product mixed state  characterized earlier. It can both certify experimentally relevant noisy resources as $2$-UDA and efficiently rule out $2$-UDA for structured many-body mixed states, by the preceding results of this work. The global mixed-state problem can often be reduced to lower-rank components or fewer-body subsystems, making the verification of the UDA property more tractable.

\section{Additivity of mixed UDA states}
\label{sec:additivity}
In this section, we analyze the additivity of bipartite  1-UDA, three-qubit 2-UDA and some $n$-partite states, which is fully characterized in Proposition \ref{pro:bipartite1-UDA} and Theorem \ref{th:rhoA1 ox A23 iff}. It is shown that the additivity of mixed UDA states can only be satisfied by imposing additional conditions on their marginals. We shall emphasize that UDA additivity of two states is of practical benefits. On the one hand, if the tensor product of two $k$-UDA states are still $k$-UDA, then the complexity for the reconstruction of the resulting states only increases linearly, instead of exponentially. On the other hand, entanglement structure of such states are deduced by their local information since it is certified that the particles of such states cannot be genuinely entangled.

Using Proposition \ref{pro:bipartite1-UDA}, we show the additivity of bipartite 1-UDA states. It is shown that such additivity can be satisfied if and only if one of the states is a pure product state. 
\begin{theorem}
\label{th:add2-partite}
Suppose $\a$ and $\b$ are two bipartite 1-UDA states of systems $(A_1A_2)$ and $(B_1B_2)$, respectively. Then $\a \ox \b$ is a 1-UDA state of the system $(A_1A_2B_1B_2)$ if and only if one of the two states $\a$ and $\b$ is a pure product state.
\end{theorem}
\begin{proof}
Let $\r_{A_1A_2B_1B_2}=\a\ox \b$.  Suppose $\s_{A_1A_2B_1B_2}$ is a state compatible with $\r$ by all 1-marginals. From Proposition \ref{pro:bipartite1-UDA}, we have 
\begin{eqnarray}
	\label{eq:alphaA1A2}
&&	\a_{A_1A_2}=\proj{\ps}_{A_{j_1}}\ox \a_{A_{j_2}},
\\	&&\b_{B_1B_2}=\proj{\eta}_{B_{k_1}}\ox \b_{B_{k_2}},
\end{eqnarray}
up to system permutations, where $j_1,j_2,k_1,k_2\in \{1,2\}$, $j_1\neq j_2$ and $k_1\neq k_2$.
Then $\s_{A_{j_1}}=\proj{\ps}$, $\s_{A_{j_2}}=\a_{A_{j_2}}$, $\s_{B_{k_1}}=\proj{\eta}$, $\s_{B_{k_2}}=\b_{B_{k_2}}$. From Lemma \ref{le:pure sep}, we obtain that
\begin{eqnarray}
	\s_{A_1A_2B_1B_2}=\proj{\ps}_{A_{j_1}}\ox \proj{\eta}_{B_{k_1}}\ox \s_{A_{j_2}B_{k_2}}.
\end{eqnarray}
The state $\s=\r$ if and only if $\s_{A_{j_2}B_{k_2}}=\a_{A_{j_2}}\ox \b_{B_{k_2}}=\s_{A_{j_2}}\ox \s_{B_{k_2}}$, i.e. $\s_{A_{j_2}B_{k_2}}$ is a 1-UDA state.
Using Proposition \ref{pro:bipartite1-UDA}, $\s_{A_{j_2}B_{k_2}}$ is 1-UDA if and only if one of $\a_{A_{j_2}}$ and $\b_{B_{k_2}}$ is pure, that is, $\a$ or $\b$ is a pure product state. 
This completes the proof.
\end{proof}

The additivity of two three-qubit product 2-UDA states is rather complicated.  
By Theorem \ref{th:rhoA1 ox A23 iff}, one can determine whether a product state $\r_{A_1}\ox \r_{A_2A_3}$ is 2-UDA directly.  We use this fact  to investigate the additivity of mixed UDA states. Let
$\a=\a_{A_1}\ox \a_{A_2A_3}$ and $\b=\b_{B_1}\ox \b_{B_2B_3}$ be two 2-UDA states. We consider the necessary and sufficient condition that $\a\ox \b$ is 2-UDA. 
Using Theorem \ref{th:rhoA1 ox A23 iff}, up to local equivalence, $\a$ and $\b$ can be assumed to be in the following form,
\begin{eqnarray}
&&\a_1=\proj{\ps_1}_{A_1}\ox \a_{A_2A_3},
\\
&& \a_2=\proj{\ps_2}_{A_2}\ox \a_{A_1}\ox \a_{A_3}
\;
\\
&&(\text{or } \a_2=\proj{\ps_3}_{A_3}\ox \a_{A_1}\ox \a_{A_2}),
\nonumber\\
&&
\a_3=\a_{A_1}\ox \proj{\eta}_{A_2A_3},
\\
&&
\a_4=\a_{A_1}\ox (H\oplus 0)_{A_2A_3}, \text{ for } \rank(\a_{A_1})=2, \;
\end{eqnarray}
where the diagonal elements of $H$ in \eqref{eq:D+0oxH+0} are positive. 
Similarly, we have
\begin{eqnarray}
&&\b_1=\proj{\xi_1}_{B_1}\ox \b_{B_2B_3},
\\
&& \b_2=\proj{\xi_2}_{B_2}\ox \b_{B_1}\ox \b_{B_3} \;
\\
&&(\text{or }\b_2=\proj{\xi_3}_{B_3}\ox \b_{B_1}\ox \b_{B_2}),
\nonumber\\
&&
\b_3=\b_{B_1}\ox \proj{\mu}_{B_2B_3},
\\
&&
\label{eq:beta4}
\b_4=\b_{B_1}\ox (D \oplus 0)_{B_2B_3}, \text{ for } \rank(\b_{B_1})=2, \;
\end{eqnarray}
with the positive diagonal elements of $D$ in \eqref{eq:D+0oxH+0}.
Suppose $\s$ is a state compatible with $\a_j\ox \b_k$ by all possible 2-marginals. It suffices to investigate the 2-UDA property of $\a_j\ox \b_k$. Without loss of generality, this derives ten cases technically, and the following analysis holds, up to swapping the $\a_{A_j}$'s with the $\b_{B_j}$'s.
Suppose $\s$ is compatible with $\a_1\ox \b_1$. From Lemma \ref{le:pure sep}, we have $\s=\proj{\ps_1}_{A_1} \ox \proj{\xi_1}_{B_1} \ox 
\s_{A_2A_3B_2B_3}$,
where $\s_{A_2A_3}=\a_{A_2A_3}$, 
$\s_{B_2B_3}=\b_{B_2B_3}$ and $\s_{A_mA_n}=\a_{A_m}\ox \b_{B_n}$. One can verify that $\a_1\ox \b_1$ is 2-UDA if and only if $\s_{A_2A_3B_2B_3}=\a_{A_2A_3}\ox \b_{B_2B_3}$. That is,  $\a_{A_2A_3}\ox \b_{B_2B_3}$ is 2-UDA and thus satisfies the conditions in Theorem \ref{th:RA12oxRA34}.
Similarly, we consider the remaining nine cases and obtain the necessary and sufficient condition under which the additivity of two three-qubit UDA states holds.
\begin{theorem}
\label{th:add3-qubit}
Suppose $\a$ and $\b$ are two three-qubit product 2-UDA states of systems $(A_1A_2A_3)$ and $(B_1B_2B_3)$, respectively. Then $\a \ox \b$ is a 2-UDA state of the system $(A_1A_2A_3B_1B_2B_3)$ if and only if one of the following conditions holds, up to swapping $\a_{A_m}$ with $\b_{B_m}$,

(i) $\a_{A_2A_3}$ and $\b_{B_2B_3}$ are pure;

(ii) $\a_{A_{j_1}}$ and $\b_{B_{2}B_3}$ (resp. $\a_{A_{j_1}}$ and $\b_{B_{k_1}}$) are pure, and the  state $\a_{A_{j_2}A_{j_3}}\ox\b_{B_1}$ (resp. $\a_{A_{j_2}A_{j_3}}\ox\b_{B_{k_2}B_{k_3}}$) is 2-UDA, satisfying Theorem \ref{th:rhoA1 ox A23 iff} (resp. Theorem \ref{th:RA12oxRA34});

(iii) $\a_{A_2A_3}$ (resp. $\a_{A_{j_1}}$) is pure, $\rank(\b_{B_2B_3})=2$ and is LU equivalent to $D\oplus 0$ in \eqref{eq:D+0oxH+0}.  The state $\a_{A_1}\ox \b_{B_1}\ox (D\oplus 0)$ (resp. $\a_{A_{j_2}A_{j_3}}\ox\b_{B_{k_2}}\ox (D\oplus 0)$) is 2-UDA satisfying Theorem \ref{th:RA12oxRA34} (resp. Theorem \ref{th:5-qubit}).

(iv) $\a_{A_1}$ and $\b_{B_1}$ are pure, $\rank(\a_{A_2A_3})=
\rank(\b_{B_2B_3})=2$, and $\a_{A_2A_3}\ox \b_{B_2B_3}$ is LU equivalent to $(H\oplus 0)\ox (D\oplus 0)$ in \eqref{eq:D+0oxH+0}. 
\end{theorem}

We recall that Theorem 1 in \cite{add2023shen} shows that if two mixed states $\a$ and $\b$ are both $k$-UDA and one of them is pure, then $\a\ox \b$ is $k$-UDA directly. Obviously, this fact is included in Theorem \ref{th:add3-qubit} (ii), where a pure $\a_{A_{j_2}A_{j_3}}\ox\b_{B_1}$ naturally satisfies the conditions in Theorem \ref{th:rhoA1 ox A23 iff}. It is also shown that the purity of $\a$ or $\b$ is a rather strict condition for mixed states since the additivity also holds for mixed UDA $\a$ and $\b$.  Theorem \ref{th:add3-qubit} implies that the rank of $\a$ and $\b$ is less than four and  the rank of $\a\ox\b$ is less than four if their UDA additivity holds. Note that $\a\ox \b$ is a six-qubit state whose rank is at most 64. Hence, UDA additivity holds only for states with relatively low rank.   

We shall consider the UDA additivity of some $n$-partite states by Proposition \ref{pro:n-qubit}. We assume that the $n$-partite states $\a_1= \ox_{i=1}^n \a_{A_i} $ and $\b_1=\ox_{j=1}^m \b_{B_j}$  be two $k$-UDA states, then at least $n-k$ and $m-k$ 1-marginals of $\a_1$ and $\b_1$ are pure, respectively. Proposition \ref{pro:n-qubit} (i) further shows that for $k_1\geq n-k$, $k_2\geq m-k$ and $k_1+k_2\geq m+n-k$, 
$\a_1\ox \b_1$ is $k$-UDA iff at least $k_1$ and $k_2$ 1-marginals of $\a_1$ and $\b_1$ are pure, respectively. Similarly, the UDA additivity of the following mixed states can be derived. The additivity of $\ox_{i=1}^n \a_{A_i}$ and  $\ox_{j=1}^{m-2}\b_{B_j}\ox \b_{B_{m-1}B_m}$ (resp. $\ox_{j=1}^{m-4}\b_{B_j}\ox \b_{B_{m-3}B_{m-2}}\ox \b_{B_{m-1}B_m}$) can be derived directly by Proposition \ref{pro:n-qubit} (ii)( resp. (iii)). Besides, the UDA additivity of  $\ox_{i=1}^{n-2} \a_{A_i}\ox \a_{A_{n-1}A_n}$ and  $\ox_{j=1}^{m-2}\b_{B_j}\ox \b_{B_{m-1}B_m}$ is also shown by Proposition \ref{pro:n-qubit} (iii).

To conclude, the preceding analysis shows that the UDA additivity cannot hold without additional conditions. The necessary and sufficient conditions of $\a$ and $\b$, which derive the additivity is given above. This answers the question we proposed at the beginning of this work. Obviously, these conditions require the two states satisfying UDA additivity to have ranks much lower than the rank threshold that guarantees each state to be UDA.

\section{Applications}
\label{sec:application}

One direct consequence of the characterizations in Sections \ref{sec:mixedUDA} and \ref{sec:additivity} is that several explicitly identified families of mixed states can be reconstructed from low-order marginals once the corresponding UDA conditions are verified. Hence, the main contribution of this section is to identify concrete low-rank families for which local measurements are sufficient.

An $n$-qubit density matrix $\r$ is parameterized by
\begin{eqnarray}
\rho=\frac{1}{2^{n}}\sum_{i_{1},\ldots,i_{n}=0}^{3}\pi_{i_{1}\cdots i_{n}}\,
\sigma_{i_{1}}^{(1)}\otimes\cdots\otimes\sigma_{i_{n}}^{(n)}. \nonumber
\end{eqnarray}
The reconstruction of $\r$ usually requires the expectation values $\pi_{i_{1}\cdots i_{n}}$ of $D_U=4^n-1$ Pauli observables and many samples of $\r$.  
The complexity is later lowered in fully quantum state tomography (FQST), where $D_F=3^n$ global Pauli observables are required by ignoring $\s_0$ in each qubit.  However, FQST is unfeasible for large systems.  By using the complete characterization of UDA states, global states can be reconstructed from measurements on $k$-marginals, a technique known as local measurements on reduced density matrices (LQST). Generally, LQST requires measuring $D_{L}=3^k \binom{n}{k}$ local observables. 
The Parallel quantum state tomography (PQST) protocol for $n$-qubit states \cite{experiment2024hu}  exploits parallel Pauli measurements to efficiently obtain all $k$-marginals, and reconstructs the global state via tensor-network learning. By choosing $k=2$, PQST requires $D_P^{(2)}=3+6\lceil \log_{2} n \rceil$ observables. 

LQST is grounded in the fact that pure states are generally UDA. However, almost all mixed states are not UDA by Corollary \ref{cor:almost}, and the complete characterization of mixed UDA states is necessary in their LQST. Besides, PQST reconstructs the global state via tensor-network learning, without assuming the UDA property. For mixed states known a priori to be 2-UDA, PQST may be implemented with a reduced set of observables, which can improve reconstruction efficiency. In fact, the set of global Pauli observables $\cM_{\text{PQST}}$ is sufficient to reconstruct $k$-marginals, which implies the observables used in PQST of mixed UDA states are less than generic states, i.e. $D_{\text{PQST}}^{\text{UDA}}\leq D_{\text{PQST}}^{\text{generic}}$. Further, a UDA state removes unidentifiable directions in the reconstruction. Statistical noise is less likely to induce large deviations, thereby reducing estimation variance.  This improves statistical efficiency and leads to higher reconstruction fidelity.
Since the measurement cost and noisy error grow with $k$, we consider the tomography of 2-UDA states in four types of tomography, and compare the number of  observables in TABLE \ref{tab:tomography}. One can see that the number of observables increases exponentially for usual tomography and FQST. Using the characterization of mixed UDA state, LQST requires much fewer observables, which grows polynomially with $n$.  Combining with the results of mixed UDA states, the observables in PQST may be less than $D_P$ in TABLE \ref{tab:tomography}.  This can substantially reduce the measurement and computational resources required for state reconstruction within the UDA families identified in this work.
\begin{table*}
    \centering
    \begin{tabular}{|c|c|c|c|c|}
\hline
 Representative state family        & $D_U$ & $D_F$ & $D_L$ & $D_P$  
\\ \hline
Three-qubit state $\r_{A_1A_2A_3}$ in Theorem \ref{th:rhoA1 ox A23 iff}       
&63  & 27 & 27 &   $15$
\\ \hline
Four-qubit state $\r_{A_1A_2}\ox \r_{A_3A_4}$ in Theorem \ref{th:RA12oxRA34}        
& 255  & 81 & 54 &   $15$
\\ \hline
Five-qubit state $\r_{A_1}\ox \r_{A_2A_3}\ox \r_{A_4A_5}$ in Theorem \ref{th:5-qubit}        
& 1023  & 243 & 90 &  $21$ 
\\ \hline
Six-qubit state $\a\ox \b$ in Theorem \ref{th:add3-qubit}        
& 4095  & 729 & 135 &  $ 21$ 
\\ \hline
\makecell{$n$-qubit state $\ox_{i=1}^{n-4} \r_{A_i}\ox \r_{A_{n-3}A_{n-2}} \ox \r_{A_{n-1}A_{n}}$\\ in Proposition \ref{pro:n-qubit}}
& $4^n-1$  & $3^n$ & $3^k \binom{n}{k}$ &  $3+6\lceil \log_{2} n \rceil$
\\\hline
    \end{tabular}
    \caption{The number of observables required in four types of quantum tomography. Here $D_U$, $D_F$, $D_L$, $D_P$ denote the number of observables used for usual tomography, FQST, LQST, PQST, respectively. }
    \label{tab:tomography}
\end{table*}

\section{Conclusion and further perspectives}
\label{sec:conclusion}
In summary, we have characterized several classes of mixed UDA states, and analyzed their additivity by giving the necessary and sufficient conditions. We have shown the structural properties of mixed UDA states. Using these facts, we show the complete characterization of mixed product states of bipartite to $n$-partite. 
We further proved rank-based obstructions showing that sufficiently high rank rules out $k$-UDA, and consequently that almost all mixed states are not $k$-UDA for even large $k$.
We established a systematic procedure for determining $k$-UDA states, and analyzed the UDA additivity of bipartite  1-UDA, three-qubit 2-UDA and some $n$-partite states. Taken together, these results show that mixed UDA property is fundamentally different from its pure-state counterpart: it is not generic, it is strongly constrained by rank and positivity, and its additivity requires additional low-rank structure.

Many problems arising from this work can be further explored. 
First, we know that a full-rank  state is not UDA, while a low-rank state may be UDA.
It is interesting to analyze the threshold of the rank of a mixed state, where states whose rank exceeds this threshold are not UDA. Second, necessary and sufficient conditions for arbitrary multipartite states remain to be derived. can be further considered. The techniques of purification of mixed states and range inclusion may be utilized. Besides, merging two systems as a local one may help to the determination of UDA states. For instance, the four-partite state $\a:=\a_A\ox \a_{BCD}$ is 2-UDA may imply that a tripartite state $\b$ is 2-UDA, where $\b$ is the same as $\a$, but regarded as a tripartite state by merging systems $A$ and $B$. 
Finally, The UDA additivity of arbitrary multipartite mixed states can be further investigated, and    
the conditions of reversible channel and Petz recovery maps \cite{jencova2012reversible} may contribute to this topic.

In addition to QST, we expect the characterization of mixed UDA states sheds light on other applications as follows. 
\begin{itemize}
  \item \textbf{Entanglement detection and distribution.}
  When an entangled state is UDA, entanglement can be certified indirectly through local measurements \cite{yu2020multi,shi2025entangle}. Otherwise, a non-$k$-UDA state implies the irreducible $(k+1)$-partite correlations \cite{irr2008zhou}.

  \item \textbf{Security in quantum communication and cryptography.}
 An adversary may attempt to distribute a counterfeit global state that reproduces the same local statistics as a desired resource state.
  If the target state is UDA with respect to the monitored data, such forgery is impossible \cite{huang2025certify,swingle2014reconstruct}.

  \item \textbf{Quantum error correction and recoverability.}
  A UDA state reflects a form of information redundancy: the global quantum state is uniquely encoded in partial subsystems.
  This is related to quantum error correction, where recovering a state from partial information is a key requirement \cite{gupta2024encode}.

  \item \textbf{Characterization of many-body quantum phases.}
  States exhibiting topological order or long-range entanglement are generally not UDA with respect to local reduced states \cite{lee2025topo}, whereas short-range entangled phases often admit UDA properties.
  This connects UDA states to the study of quantum phases and phase classification \cite{will2025probing}.
\end{itemize}

\appendices
\makeatletter
\@addtoreset{definition}{section}
\makeatother
\setcounter{definition}{0}
\renewcommand{\thedefinition}{\Alph{section}.\arabic{definition}}

\section{Proofs of general properties of $k$-UDA states}
  \label{app:lemma}
In addition to the equation of  $k$-marginals, the positive semi-definiteness  is a natural restriction of a compatible state.  To prove that some states are not $k$-UDA, we construct another state compatible with them, and the semi-definiteness of these states is guaranteed by the following fact. 
\begin{lemma}
	\label{le:Y-ve x}
	If $X$ and $Y$ are two positive semidefinite matrices, and $\cR(X) \subseteq \cR(Y)$, then there is a small enough positive number $\ve>0$ such that $Y - \ve X \geq 0$. 
\end{lemma}
\begin{proof}
	From $X\geq 0$ and $Y\geq 0$, there is an invertible matrix $P$ such that $P^\dg X P=\bma D_1 &0\\ 0&0\ema$ and 
	$P^\dg Y P=\bma D_2 &0\\ 0&0\ema$, where $D_1=\diag\{s_1, s_2, ..., s_{r_1}\}$ and $D_2=\diag\{t_1, t_2, ..., t_{r_2}\}$ are diagonal matrices with positive elements.
	By $\cR(X) \subseteq \cR(Y)$, we have $\rank(D_1)\leq \rank(D_2)$. Let $\ve =\min_{k=1}^{r_1} \frac{t_k}{s_k}$. Then $P^\dg (Y - \ve X) P\geq 0$, i.e. $Y - \ve X\geq 0$.  
\end{proof}
Although we may establish different states by Lemma \ref{le:Y-ve x}, we shall emphasize that the positive semi-definiteness is a power restriction  that helps to eliminate the uncertainty of the state and conclude the UDA property.

Lemma \ref{le:lr+(1-l)s_not2uda} shows five properties of $k$-UDA states. First, $\r\ox \s$ is not $k$-UDA if either of $\r$ and $\s$ is not $k$-UDA.
Second,
a state whose range includes a non-$k$-UDA state is itself not $k$-UDA. Third, two states that have the same range are $k$-UDA or not at the same time. Fourth, two states $\a\ox \b$ and $\b\ox \a$ are the same time $k$-UDA or not.
Fifth,  if two states are $k$-compatible, then their marginals are also $k$-compatible.
The proof is given as follows. 

\begin{proof}
(i) We denote by $A_{[m]}$ and $B_{[n]}$ the $m$-partite system $(A_1,...,A_m)$ and $n$-partite system $(B_1,...,B_n)$, respectively. 
Suppose $\r_{A_{[m]}}$ is not $k$-UDA and $\s_{B_{[n]}}$ is arbitrary. Then there is another state $\g_{A_{[m]}}$ that is $k$-compatible with $\r$, i.e. $\r_\cS=\g_\cS$ for $\abs{\cS}=k$. We show that $\g\ox\s$ is $k$-compatible with $\r\ox \s$. Let  $\cS=(S_1,S_2)$ and $S_1\in A_{[m]}$, $S_2\in B_{[n]}$.
\begin{eqnarray}
(\g\ox \s)_\cS&=&\tr_{\{A_{[m]}\backslash S_1\}\cup \{B_{[n]}\backslash S_2\}} (\g\ox \s)
\\	\nonumber
&=&
\g_{S_1}\ox \s_{S_2}
=(\r\ox \s)_\cS,
\nonumber
\end{eqnarray}
where $\r_{S_1}=\g_{S_1}$ by $S_1\subseteq\cS$. 
From $\r\neq \g$, we have $\r\ox \s\neq \g\ox \s$.
Hence, $\r\ox \s$ is not $k$-UDA.

(ii) Suppose $\r$ and $\s$ are on $\cH_{A_1A_2...A_n}$.  Since $\r$ is not $k$-UDA, there is a state $\r'$ such that $\r_\cS=\r'_{\cS}$ for all $\cS\subset(A_1A_2...A_n)$ and $\abs{\cS}=k$. Then $\l\r_\cS+(1-\l)\s_\cS=\l\r'_\cS+(1-\l)\s_\cS$ implies that $\l\r'+(1-\l)\s$ is $k$-compatible with $\l\r+(1-\l)\s$, and they are different states. 

(iii) From $\cR(\b)\subset \cR(\a)$, we have $\a=(\a-\e\b)+\e\b$ for $\e>0$ small enough. Using Lemma \ref{le:Y-ve x}, we have $\a-\e\b\geq 0$ is a (unnormalized) state. From Lemma \ref{le:lr+(1-l)s_not2uda}, if $\b$ is not $k$-UDA, then $\a$ is not $k$-UDA. Similarly, by $\cR(\a)\subset \cR(\b)$, $\a$ is not $k$-UDA implies that $\b$ is not $k$-UDA. 

(iv) Let $\r$ and $\s$ be two states of systems $(A_1A_2...A_m)$ and $(B_1B_2...B_n)$, respectively. Suppose $\r\ox \s$ is not $k$-UDA. There is a different $(m+n)$-partite  state $\g_{A_{[m]}B_{[n]}}$ that is $k$-compatible with $\r\ox \s$, i.e. $\g_{A_\cK B_\cP}=\r_{A_\cK} \ox \s_{B_\cP}$ with $\abs{A_\cK}+\abs{B_\cP}=k$. 
Obviously, $\g_{B_\cP A_\cK }=\s_{B_\cP}\ox \r_{A_\cK}$. That is,  the $k$-partite marginals of the state $\g_{B_{[n]}A_{[m]}}$ is the same as that of $\s\ox \r$. Since $\g_{A_{[m]}B_{[n]}}\neq \r\ox \s$, we have $\g_{B_{[n]}A_{[m]}}\neq \s\ox \r$. 

(v) Since  $\r$ and $\s$ are $k$-compatible, we can trace out the systems $\cS^c$ on $\r$ and $\s$, and the resulting states are also $k$-compatible. 
This completes the proof.
\end{proof}

Lemma \ref{le:general_properties} shows the general properties of $k$-UDA states and of pairs of 
$k$-compatible states by studying their relations through the ranges of the corresponding density operators. 
The proof is shown here. 

\begin{proof}
(i) Let $\cS\subset(A_1A_2...A_n)$ and $\abs{\cS}=k$. For $\r=\text{Re} (\r)+i \text{Im}(\r)$, we have $\text{Re}(\r)_\cS=\frac{1}{2}(\r_\cS+\r^*_\cS)$. Since $\r$ is $k$-compatible with $\s$, we have $\r_\cS=\s_\cS$ and $\r^*_\cS=\s^*_\cS$. Then $\text{Re}(\r)_\cS=\frac{1}{2}(\s_\cS+\s^*_\cS)=\text{Re}(\s)_\cS$. Hence $\r$ has the same real parts as $\s$. The equivalence of imaginary parts $\text{Im} (\r)$ and $\text{Im} (\s)$ can be shown similarly. 

(ii) We show the first fact as follows.  Suppose $A:=\text{Re}(\r)=\frac{\r+\r^*}{2}$ and $B:=\text{Im}(\r)=\frac{\r-\r^*}{2i}$, which are symmetric and antisymmetric, respectively. Let $z=x+iy\in \bbC^n$, for $x,y\in \bbR^n$. We have $z^\dg\r z=\bma x^T & y^T\ema M \bma x\\y \ema\geq 0$, for $M=\bma A&B\\-B&A\ema$.   That is, $M\geq 0$. For any $u=x+iy\in \ker (A)$, we have $Ax=Ay=0$. By choosing $v_1=(x,0)^T$ and $v_2=(0,y)^T$, we have $Bx=By=0$, i.e. $u\in \ker(B)$. Then $\cR(B^\dg)\subseteq\cR(A^\dg)$, i.e. $\cR(B)\subseteq\cR(A)$. 
Next we show the second fact. Suppose $\text{Im}(\r)\neq 0$. We have  $\r^*\neq \r$. Since all $k$-marginals of $\r$ are real, we have $\r_\cS=\r_{\cS}^*$ with $\abs{\cS}=k$. That is, $\r^*$ is another state $k$-compatible with $\r$. 

(iii) It suffices to show that $\ker (\r_\cS)\subseteq\ker(H_\cS)$. For any $\ket{x}\in \ker(\r_\cS)$, we have $\r_\cS\ket{x}=0$, which is equivalent to $\bra{x}\r_\cS\ket{x}=0$. Since $\bra{x}\r_\cS\ket{x}=\sum_j\bra{x,j}\r\ket{x,j}= 0$, we have $\r\ket{x,j}=0$, i.e. $\ket{x,j}\in \ker(\r)$. By $\ker(\r)\subseteq \ker(H)$, we have $\ket{x,j}\in \ker(H)$. We show that $\ket{x}\in \ker H_\cS$. For any $\ket{u}\in \cH_\cS$, we have $\bra{x}H_S\ket{u}=\sum_j\bra{x,j}H\ket{u,j}=0$. Hence $H_\cS\ket{x}=0$, i.e. $\ket{x}\in \ker (H_\cS)$. 
The converse does not generally hold, for example, $H=\ketbra{01}{01}$ and $\r=\proj{\ps}$ with $\ket{\ps}=\ket{00}+\ket{11}$. 

(iv) Let $\cS\subset[A_1A_2...A_n]$ and $\abs{\cS}=k$. If $A_1\notin \cS$, we have $[(\L_{A_1}\ox I)\eta]_\cS=\eta_\cS$ for $\eta=\a,\; \b$.  By $\a_\cS=\b_\cS$, we prove the claim. Otherwise, we have $A_1\in \cS$ and $[(\L_{A_1}\ox I)\eta]_\cS=(\L_{A_1}\ox I_{\cS\backslash A_1})\eta_\cS$. Hence $[(\L_{A_1}\ox I)\a]_\cS=[(\L_{A_1}\ox I)\b]_\cS$, and $(\L_{A_1} \otimes I_B) \alpha$ and $(\L_{A_1} \otimes I_B) \b$ are $k$-compatible. Further, if $\L_{A_1}$ is reversible, there is another channel $R_A$, such that $[(R_A\circ \L_A)\ox I]\a=\a$. If $(\L_A\ox I)\a$ is not $k$-UDA, then there is another state $\b$  $k$-compatible with it. By the preceding analysis, $(R_A\ox I)\b$ is  $k$-compatible with $\b$, and thus compatible with $\a$. Here  $(R_A\ox I)\b\neq \a$ by $(\L_A\ox I)\a\neq \b$, which derives that $\a$ is not $k$-UDA.  Similarly,  $\a$ is not $k$-UDA derives that $(\L_A\ox I)\a$ is not $k$-UDA. 

(v) Suppose $\cR(\r_{A_1A_2...A_n})=\text{span}\{\ket{\a_k}: k=1,2,...,r\}$.
Let $\ket{\ps}$ be the purification of $\rho_{A_1A_2...A_n}$ on systems $A_1A_2...A_{m}$. Then 
$$\ket{\ps}=
\sum_k \ket{\a_k}_{A_1...A_n}\ket{\b_k}_{A_{n+1}...A_{m}}.$$
 For any $n$-partite channel $\L(\cdot)=\sum_{j=1}^k M_j(\cdot) M_j^\dg$ acting on $\cH_{A_{n+1}...A_{m}}$, we have
\begin{eqnarray}
\s=&&\bbI_{A_1A_2...A_n}\ox \L_{A_{n+1}A_{n+2}...A_{2n}}(\proj{\ps})
\\
=&& \sum_j  (\bbI\ox M_j) \proj{\ps} (\bbI\ox M_j^\dg). 
\nonumber
\end{eqnarray}
Since $\r_{A_{n+1}...A_m}$ is $n$-UDA, we have $\s_{A_{n+1}...A_{m}}=\r_{A_{n+1}...A_{m}}$. Hence, $ M_j\ket{\b_k}\in \cR(\rho_{A_{n+1}...A_{m}})$. Hence $\cR(\s)\subseteq \cR(\rho)$.    
\end{proof}

\section{Auxiliary results for  tripartite mixed UDA states}
\label{app:3-qubit}

The UDA property is invariant up to local unitaries. It allows us to simplify a bipartite state as follows. 
\begin{lemma}
	\label{le:simplify2-qubit}
	Let $\eta$ be a bipartite  state supported on $\bbC^{m}\ox \bbC^n$ and $\rank (\eta)=r$ with $1\leq r\leq m+n-2$. Up to local unitaries, the last diagonal element of $\eta$ is zero, i.e. $\eta=M\oplus0$ with $M\geq 0$ and $\tr(M)=1$.
\end{lemma}
\begin{proof}
The dimension $d$ of the kernel of $\eta$ satisfies that  $d=mn-r\geq (m-1)(n-1)+1$. 	By the theory of Segre variety,   there is a product state in the kernel of $\eta$. Up to local unitaries, we assume that this product is $\ket{m-1,n-1}$. Then the range of $\eta$ does not contain $\ket{m-1,n-1}$, and thus $\eta=M\oplus 0$.
\end{proof}

By choosing $m, n, r=2$ in Lemma \ref{le:simplify2-qubit}, we  show the necessary and sufficient conditions under which  a set of product mixed states $\cP=\{\r_{A_1}\ox \r_{A_2A_3}: \rank(\r_{A_2A_3})=2 \text{ or }\rank(\r_{A_2A_3})=3 \text{ with a zero diagonal element}\}$ is 2-UDA as follows.
\begin{proposition}
\label{pro:D+0iff|D|=0}
	Suppose a three-qubit state $\r=\r_{A_1}\ox \r_{A_2A_3}$ with 
	\begin{eqnarray}
		\label{eq:rhoA23=D+0}
		\r_{A_2A_3}= D\oplus 0 \text{ for } D=\bma q_1 &y & z\\
		y^* &q_2&x \\
		z^*& x^* &q_3 \ema. 
	\end{eqnarray}
	Then 
	(i) if $\rank(\r_{A_1})=1$ or $q_2=0$ or
	$q_3=0$, then $\r$ is 2-UDA.
	
	(ii) if $\rank(\r_{A_1})=2$ and $q_k>0$, then  $\r$ is 2-UDA iff  
	$\abs{D}=0$.
	
	(iii) if $\rank(\r_{A_1})=2$ and  $q_1=0$, then $\r$ is 2-UDA iff $\abs{x}=\sqrt{q_2q_3}$. 
\end{proposition}
\begin{proof}
	(i) If $\rank(\r_{A_1})=1$ or $q_2=0$ or $q_3=0$, then one of a single-qubit marginals of $\r$ is pure. Hence $\r$ is 2-UDA.  
	
	(ii) Up to local unitaries, we assume that $\r_{A_1}=p\proj{0}+(1-p)\proj{1}$. 
	We show the "if" part. 
	Let $\sigma$ be the state compatible with $\r$. Since $\s$ equal to all 2-marginals of $\r$, we obtain that $\s$ is given by
	\begin{eqnarray}
		\bma
		pq_1&py&pz&0&0&0&0&0\\
		py^*&pq_2&s_{23}&0&0&0&s_{27}&0\\
		pz^*&s_{23}^*&pq_3&0&0&s_{36}&0&0\\
		0&0&0&0&0&0&0&0\\
		0&0&0&0&(1-p)q_1&(1-p)y&(1-p)z&0\\
		0&0&s_{36}^*&0&(1-p)y^*&(1-p)q_2&x-s_{23}&0\\
		0&s_{27}^*&0&0&(1-p)z^*&x^*-s_{23}^*&(1-p)q_3&0\\
		0&0&0&0&0&0&0&0\ema.
        \nonumber
	\end{eqnarray} 
	From $\abs{D}=0$,  
	there is a nonzero vector $v=[a,b,c]^\dg$ such that $Dv=0$. That is, 
	\begin{eqnarray}
		\label{eq:Dv=0}
		\begin{cases}
			q_1a+yb+zc=0\\
			y^*a+q_2b+xc=0\\
			z^*a+x^*b+q_3c=0.
		\end{cases}   
	\end{eqnarray}
	Suppose $B_1$ and $B_2$ are the three-order principle minors formed from rows and columns (1, 2, 3) and  (5, 6, 7) of the $\s$, respectively.
	Note that $B_1+B_2=D$.
	We have $B_1v+B_2v=0$, and hence $v^*B_1v+v^*B_2v=0$. 
	From  $B_1,B_2\geq 0$, we have $v^*B_1 v\geq 0$ and $v^*B_2 v\geq0$. Then  $v^*B_1 v=v^*B_2 v=0$, and thus $B_1v=B_2v=0$.
	Using \eqref{eq:Dv=0}, we have
	\begin{eqnarray}
		\label{eq:B_1v=-B_2v}
		B_1v=-B_2v=
		[0,c(s_{23}-px),b(s_{23}^*-px^*)]^\dg=0.
	\end{eqnarray}
	From the first equality of \eqref{eq:Dv=0},  $b=c=0$ implies that $a=0$ by $q_1>0$, and thus $v=0$, which is a contradiction.
	It holds that $c\neq 0$ or $b\neq 0$ in $v$. By \eqref{eq:B_1v=-B_2v}, we have $s_{23}=px$.

	Next we show that $s_{36}=s_{27}=0$. The positive semidefiniteness  of $\s$ implies that $\abs{x}\leq\sqrt{q_2q_3}$,   $\abs{y}\leq\sqrt{q_1q_2}$ and  $\abs{z}\leq \sqrt{q_1q_3}$.  This derives four cases, depending on the value of $\abs{x}, \abs{y}$ and $\abs{z}$.
	
	Case 1: $\abs{x}<\sqrt{q_2q_3}$,  $\abs{y}<\sqrt{q_1q_2}$ and $\abs{z}<\sqrt{q_1q_3}$ hold simultaneously. From
    \begin{eqnarray}
     &&C_1:=\bma
	pq_1&py&pz&0\\
	py^*&pq_2&px&s_{27}\\
	pz^*&px^*&pq_3&0\\
	0&s_{27}^*&0&(1-p)q_3\ema\geq 0,
\\  &&C_2:=\bma
	pq_1&py&pz&0\\
	py^*&pq_2&px&0\\
	pz^*&px^*&pq_3&s_{36}\\
	0&0&s_{36}^*&(1-p)q_2\ema\geq 0,
    \end{eqnarray}
	we have 
	$(1-p)p^3q_3\abs{D}-\abs{s_{27}}^2p^2(q_1q_3-\abs{z}^2)\geq 0$,
	$ (1-p)p^3q_2\abs{D}-\abs{s_{36}}^2p^2(q_1q_2-\abs{y}^2)\geq 0$.
	By $\abs{D}=0$, $\abs{y}<\sqrt{q_1q_2}$ and $\abs{z}<\sqrt{q_1q_3}$, we obtain that $s_{27}=s_{36}=0$.

	Case 2: $\abs{x}=\sqrt{q_2q_3}$.
	By the positive semi-definiteness of $\s$, the following matrices are positive semidefinite
    \begin{eqnarray}
    \bma
	pq_2&px&s_{27}\\
	px^*&pq_3&0\\
	s_{27}^*&0&(1-p)q_2\ema, 
    \quad
    \bma
	pq_2&px&0\\
	px^*&pq_3&s_{36}\\
	0&s_{36}^*&(1-p)q_2\ema
    \nonumber
    \end{eqnarray}
 We have 
		$p^2(1-p)q_2(q_2q_3-\abs{x}^2)-pq_3\abs{s_{27}}^2=-pq_3\abs{s_{27}}^2\geq0$,
	and $s_{27}=0$ by $pq_3>0$.
	Similarly, $D_2\geq 0$ derives that $s_{36}=0$. 
	
	Case 3: $\abs{y}=\sqrt{q_1q_2}$.
	We have  $s_{27}=s_{36}=0$ by the positivity of the following matrices,
    \begin{eqnarray}
   \bma
	pq_1&py&0\\
	py^*&pq_2&s_{27}\\
	0&s_{27}^*&(1-p)q_3\ema,
    \quad
   \bma
	pq_3&0&s_{36}\\
	0&(1-p)q_1&(1-p)y\\
	s_{36}^*&(1-p)y^*&(1-p)q_2\ema.\nonumber
    \end{eqnarray}

	Case 4: $\abs{z}=\sqrt{q_1q_3}$. Then  $s_{27}=s_{36}=0$ by the positivity of the following matrices,
\begin{eqnarray}
\bma
	pq_1&pz&0\\
	pz^*&pq_3&s_{36}\\
	0&s_{36}^*&(1-p)q_2 
	\ema,  \quad
    \bma
	pq_2&0&s_{27}\\
	0&(1-p)q_1&(1-p)z\\
	s_{27}^*&(1-p)z^*&(1-p)q_3\ema. \nonumber  
\end{eqnarray}	
	To conclude, we show that $s_{23}=px$ and $s_{27}=s_{36}=0$, i.e. $\s=\r$, by assuming that $\abs{D}=0$. 
	We have proven the "if" part.

	We show the "only if" part. 
	Suppose $\abs{D}>0$. Then
	$\abs{x} < \sqrt{q_2 q_3}$, $\abs{y}<\sqrt{q_1q_2}$ and $\abs{z}<\sqrt{q_1q_3}$ hold simultaneously. 
	Let $h_1=q_2q_3-\abs{x}^2$, $h_2=q_1q_2-\abs{y}^2$ and $h_3=q_1q_3-\abs{z}^2$. Then $h_k>0$.
	We show that there is another state $\s_1$ compatible with $\r$ by all possible 2-marginals.
	In detail, we set all elements of $\s_1$ the same as $\r$, except for $s_{27}$. Now we show that the element $s_{27}$ of $\s_1$ is not necessarily equal to zero.  
	We check all principal minors of $\s_1$ that contains the element $s_{27}$, and obtain that
	\begin{eqnarray}
		\label{eq:rangeS27}
	0<\abs{s_{27}}^2\leq f,
	\end{eqnarray}
	where $f=\min\{(1-p)p h_2h_3/q_1^2,(1-p)p h_1 h_2/q_1q_2,
		p(1-p)h_1 h_3/q_1q_3, p(1-p)h_1^2/q_2q_3, 
		p(1-p)h_1t/q_3h_2,p(1-p)t/q_1,
        (1-p)p h_1 t/q_2h_3,
		p(1-p)t^2/h_2h_3\}$ with  $t=\abs{D}>0$. 
	Note that the RHS of \eqref{eq:rangeS27} is strictly larger than zero. Hence we can choose $s_{27}>0$ satisfying \eqref{eq:rangeS27} in $\s_1$.
	One can verify that $\sigma_1$ is compatible with $\r$ by all possible 2-marginals and $\sigma_1\neq \r$.
	Hence $\r$ is not 2-UDA.
	This completes the proof. 
	
	(iii)  We show the "if" part. Let $\s$ be the state compatible with $\r$. Since $\s$ equal to all 2-marginals of $\r$, we obtain that
	\begin{eqnarray}
    \label{eq:s=0+H+0}
		\s=0\oplus \bma
	pq_2&s_{23}&0&0&0&s_{27}\\
		s_{23}^*&pq_3&0&0&s_{36}&0\\
		0&0&0&0&0&0\\
		0&0&0&0&0&0\\
		0&s_{36}^*&0&0&(1-p)q_2&x-s_{23}\\
		s_{27}^*&0&0&0&x^*-s_{23}^*&(1-p)q_3\ema \oplus 0.
	\end{eqnarray}
Let\begin{eqnarray}
		\label{B3,B4}
		&&D_{1}:=\bma
		q_2&x\\
		x^*&q_3\ema,\quad  
        B_3:=\bma
		pq_2&s_{23}\\
		s_{23}^*&pq_3\ema\geq 0, 
  \\  
 && B_4:=\bma
		(1-p)q_2&-s_{23}+x\\
		(-s_{23}+x)^*&(1-p)q_3\ema\geq 0.
	\end{eqnarray}
	Note that $B_3+B_4=D_{1}$  for $B_3$,$B_4$ in \eqref{B3,B4}. 
	From $\abs{D_1}=0$, there is a nonzero vector $v=[a,b]^\dg$ such that $D_1v=0$. That is,
	\begin{eqnarray}
		\label{eq:D_1v=0}
			q_2a+xb=0, \quad
			x^*a+q_3b=0.  
	\end{eqnarray}
	Then we have $B_3v+B_4v=0$, and hence $v^*B_3v+v^*B_4v=0$. By $B_3,B_4\geq 0$, we have $v^*B_3v\geq 0$ and $v^*B_4v\geq 0$. Then $v^*B_3v=v^*B_4v=0$, and thus $B_3v=B_4v=0$. Using \eqref{eq:D_1v=0}, we have 
	\begin{eqnarray}
		\label{B_3v=-B_4v}
		B_3v=-B_4v=[b(s_{23}-px),a(s_{23}^*-px^*)]^\dg =0
	\end{eqnarray}
	From the first equality of \eqref{eq:D_1v=0}, we have $a\neq 0$ or $b\neq 0$ in $v$. By \eqref{B_3v=-B_4v}, we have $s_{23}=px$.
	Next we show $s_{27}=s_{36}=0$.
	Since the following principal minors are positive definite, i.e.
    \begin{eqnarray}
     \bma
	pq_2&s_{23}&s_{27}\\
	s_{23}^*&pq_3&0\\
	s_{27}^*&0&(1-p)q_3\ema, \quad 
    \bma
	pq_2&s_{23}&0\\
	s_{23}^*&pq_3&s_{36}\\
	0&s_{36}^*&(1-p)q_2\ema
    \end{eqnarray}
 we have 
$(1-p)q_3\abs{D_1}-\abs{s_{27}}^2pq_3\geq 0$, 
$(1-p)q_2\abs{D_1}-\abs{s_{36}}^2pq_2\geq 0$.
By $\abs{D_1}=0$, we obtain that $s_{27}=s_{36}=0$.
	To conclude, we show that $s_{23} = px$ and $s_{27} = s_{36} = 0$, i.e. $\s =\r$, if  $\abs{x}=\sqrt{q_2q_3}$. We have proven the
	”if” part.
	
	We show the "only if" part. Suppose $\abs{D_1}>0$. We show that there is another state $\s_1$ compatible with $\r$ by all possible 2-marginals. In detail,we set all elements of $\s_1$ the same as $\r$, except for $s_{27}$. Now we show that the element $s_{27}$ of $\s_1$ is not necessarily equal to zero. We check all principal minors of $\s_1$ that contains the element $s_{27}$, and obtain that
	\begin{eqnarray}
		\label{s_{27}}
		0<\abs{s_{27}}^2\leq p(1-p)(q_2q_3-\abs{x}^2)^2/q_2q_3. 
	\end{eqnarray}
	Note that the RHS of \eqref{s_{27}} is strictly larger than zero. Hence we can choose $s_{27}>0$ satisfying \eqref{s_{27}} in $\s_1$. One can verify that $\s_1$ is compatible with $\r$ by all possible 2-marginals and $\s_1\neq \r$. Hence $\r$ is not 2-UDA. This completes the proof.
\end{proof}

Proposition \ref{pro:D+0iff|D|=0} shows that the state $\r=\r_{A_1}\ox \r_{A_2A_3}$ with a rank-three $\r_{A_2A_3}$ which has a zero diagonal element is not 2-UDA.  
The following fact is given for a high-dimensional state. By this fact,  one can obtain that $\r$ is not 2-UDA if $\rank(\r_{A_1})=2$ and $\rank (\r_{A_2A_3})=3$ or 4, by choosing $d_1=d_2=d_3=2$. 
\begin{proposition}
\label{pro:rankA23=3or 4}
	Let  $\r=\r_{A_1}\ox \r_{A_2A_3}$ be a  tripartite state supported on $\bbC^{d_1}\ox \bbC^{2} \ox \bbC^{2}$. Suppose  $\rank(\r_{A_1})=d_1$. Then 
	
	(i) If $\rank(\r_{A_2 A_3})=3$ and $\r_{A_2A_3}$ is not LU equivalent to the state with a zero diagonal element, then
	$\r$ is not 2-UDA.
	
	(ii) If $\rank(\r_{A_2 A_3})=4$, then $\r$ is not 2-UDA. 
\end{proposition}
\begin{proof}
	(i) First we claim that 
	\begin{eqnarray}
\ker(\r_{A_2A_3})
=\text{span} \{\cos\t\ket{00}+\sin\t\ket{11}, \; \t\in (0,\frac{\pi}{4}]\}.  \nonumber
	\end{eqnarray}
	In fact, we have $\dim \ker(\r_{A_2A_3})=1$. Suppose a state $\ket{\b}\in \ker(\r_{A_2A_3})$. If $\ket{\b}$ is a product pure state, then 
	$\ket{\b}$ is LU equivalent to $\ket{00}$, which implies that $\r_{A_2A_3}$ has a zero diagonal element. This shows the claim. Hence, the range of $\r_{A_2A_3}$ is given by
	\begin{eqnarray}
		\cR(\r_{A_2A_3})=\text{span}\{
		\sin \t\ket{00}-\cos\t\ket{11}, 
		\; \ket{jk}: jk\neq 00, 11\}. 
		\nonumber
	\end{eqnarray}
Suppose $\r_{A_1}=\diag\{p_1,p_2, ..., p_{d_1}\}$. 	Let
	\begin{eqnarray}
		\label{eq:sigmaA123B}
		\s_{A_1A_2A_3}=\bma p_1\r_{A_2A_3} &0 &\cdots &0 & B\\ 
		0  & p_2\r_{A_2A_3} & \cdots & 0 &0 \\
		\vdots & \vdots & & \vdots &\vdots\\
		B^\dg&0&\cdots & 0 & p_{d_1}\r_{A_2A_3}
		 \ema,
	\end{eqnarray}
	where $B$ is a matrix where only the bottom-left element is $x$, and $x>0$ small enough. 
	One can verify that the state $\s$ is compatible with $\r$ by all 2-marginals. We show that $\s\geq 0$.  From $\cR(B)\subset \cR(\r_{A_2A_3})$, the state $\s$ is congruent to the matrix
	$$ (p_1\r_{A_2A_3}-BV^\dg VB^\dg) \oplus_{k=2}^{d_1-1} p_2\r_{A_2A_3}  \oplus I_{d_2d_3-1}\oplus 0. $$
	It suffices to show that 
	\begin{eqnarray}
		\label{eq:prA23-BV}
	\xi&:=&p\r_{A_2A_3}-BV^\dg VB^\dg
   \nonumber \\
   &=& p\r_{A_2A_3}-\diag \{0,0, \cdots, 0,y\}\geq 0,  \nonumber 
	\end{eqnarray}
	where $y\ra 0$ when $x\ra 0$. By Lemma \ref{le:Y-ve x}, there is $x>0$ small enough such that $\xi \geq 0$, and hence $\s\geq 0$. This completes the proof.

	(ii) If $\rank (\r_{A_2A_3})=4$, then the state $\s$ in \eqref{eq:sigmaA123B} is also compatible with $\rho$. Further $\s\geq 0$ holds since \eqref{eq:prA23-BV} is derived by 
	$\cR(\diag \{0,0, \cdots, 0,y\})\subseteq\cR(\r_{A_2A_3})$ and Lemma \ref{le:Y-ve x}. 
\end{proof}

In Proposition \ref{pro:rho2-udaDiagonald-dim}, we show that $\r=\r_{A_1}\ox \r_{A_2A_3}$ with diagonal $\r_{A_2A_3}$ is 2-UDA if and only if  one of the six 1- and 2-marginals of $\r$ is pure. We show the proof of Proposition \ref{pro:rho2-udaDiagonald-dim} as follows.

\begin{proof}
	The if part is trivial by Lemma \ref{le:pure sep}. It remains to show the "only if" part. Suppose the  tripartite  state $\r$ is supported on $\bbC^{d_1} \ox  \bbC^{d_2} \ox \bbC^{d_3}$.
	Since the UDA property is invariant up to local unitaries, we assume that
	\begin{eqnarray}
		\label{eq:rho=p0 ox rA23}
		\r:=&&\diag\{p_1,p_2,...,p_{d_1}\}_{A_1}\ox \r_{A_2A_3}
		\\
		=&& \diag\{p_1,p_2,...,p_{d_1}\}_{A_1}\ox\diag\{M_{1}, M_{2}, \cdots, M_{d_2}\},
        \nonumber
	\end{eqnarray} 
	where
	$	p_j\in [0,1]$ and  $ 
M_k:=\diag\{q_{(k-1)d_3+1},...,q_{kd_3}\}$.
	If none of the 1- and 2-marginals of $\r$ is pure, then
	\begin{eqnarray}
		\label{eq:rA1>1}
	&& \hspace{-0.6cm}
\rank(\r_{A_1})=\rank(\diag\{p_1,p_2,...,p_{d_1}\})>1,
	\\
		&& 		\hspace{-0.6cm}\label{eq:rA2>1} 
		\rank(\r_{A_2})= \rank(\diag\{\tr(M_1),  ..., \tr(M_{d_2}) \})>1,
		\\
		\label{eq:rA3>1}
		&& \hspace{-0.6cm}
        \rank(\r_{A_3})= \rank(\sum_{k=1}^{d_2} M_k)>1.
	\end{eqnarray}		
	since all the preceding 1-marginals of $\r$ are not pure implies that all the 2-marginals $\r_{A_1A_2}$, $\r_{A_1A_3}$, $\r_{A_2A_3}$ are not pure.

	From (\ref{eq:rA1>1})-(\ref{eq:rA3>1}), at least two of $p_j$'s and $q_k$'s  are strictly larger than zero. Suppose $p_{j_1}, \; p_{j_2}>0$ for  $1\leq j_1<j_2\leq d_1$  and $q_{k_1},\; q_{k_2}>0$ with  $1\leq k_1<k_2\leq d_2d_3$. 
	Then we have a different state $\s$ compatible with $\r$, that is, 
	\begin{eqnarray}
\label{eq:defdiffs}
\s&=&\oplus_{l=1}^{j_1-1}p_l \r_{A_2A_3}
	\oplus  T_1 \oplus_{l=j_1+1}^{j_2-1}p_l \r_{A_2A_3}
    \\
&&\oplus T_2 \oplus_{l=j_2+1}^{d_1}p_l \r_{A_2A_3},\nonumber
	\end{eqnarray} 
	where the diagonal elements of $T_l$ and  are the in turns $p_{j_l}q_1$,..., $p_{j_l}q_{k_1}$, ..., $p_{j_l}q_{k_2}$, ..., $p_{j_l}q_{d_2d_3}$. The only nonzero off-diagonal elements of $T_1$ and $T_2$ are at $k_1$-th row and $k_2$-th column (and also $k_2$-th row and $k_1$-th column by Hermitian property), it is $x$ and $-x$, respectively.  
	For  the nonzero element $x$, it satisfies that
	\begin{eqnarray}
		\label{eq:rangex1}
		0<\abs{x}\leq  \min\{p_{j_1}\sqrt{q_{k_1}q_{k_2}}, \; 
		p_{j_2}\sqrt{q_{k_1}q_{k_2}} \}
	\end{eqnarray}
	The codition (\ref{eq:rangex1}) derives that $\s\geq 0$ and $\s\neq \r$. Hence $\r$ is not 2-UDA.
	This completes the proof.
\end{proof}

We turn to analyze the 2-UDA property of arbitrary three-qubit states. Proposition \ref{pro:3qubit_general} shows that if  the rank of $\rho$ is six with additional conditions, or greater than seven, then $\rho$ is necessarily not 2-UDA. The proof is presented as follows.

\begin{proof}
We show the proof of the first fact. One can show that $\cR(\r)=\cR(\r_{A_1}\ox \r_{A_2A_3})$ with full-rank $\r_{A_1}$. Let $\ket{\ps}\in \ker(\cR(\r_{A_2A_3}))$ and $\Pi=I_{A_1}\ox \proj{\ps}$. Then $0=\bra{\ps}\r_{A_2A_3}\ket{\ps}=\tr[(I_{A_1}\ox \proj{\ps})\r]=\tr(\Pi \r)$. Since $\Pi \r\Pi\geq 0$ and $\tr(\Pi \r \Pi)=\tr(\Pi \r)=0$, we have $\Pi \r\Pi=0$. This implies that   $\cR(\r)=\supp(\r)\subseteq \cH_{A_1}\ox \cR(\r_{A_2A_3})=\cR(\r_{A_1}\ox \r_{A_2A_3})$. From $\dim(\cR(\r))=\dim(\cR(\r_{A_1}\ox \r_{A_2A_3}))$, we have $\cR(\r)=\cR(\r_{A_1}\ox \r_{A_2A_3})$. Since $\r_{A_1}\ox \r_{A_2A_3}$ is not 2-UDA, we have $\r$ is not 2-UDA. 
 
 The second fact is proved by showing the range of $\r$ has a pure state $\ket{\g}$ LU equivalent to the quasi-GHZ state $a\ket{000}+b\ket{111}$ with $ab\ne0$.  More details will be given in the proof of Proposition \ref{pro:rankn-1}.
 This completes the proof. 
 \end{proof}

\section{Auxiliary results for four-qubit mixed UDA states}
\label{app:4-qubit}

First, we consider the diagonal 2-UDA state $\r_{A_1A_2} \ox \r_{A_3A_4}$ and present its fully characterization. 
\begin{proposition}
\label{pro:diag4-qubit}
Let $\r=\r_{A_1A_2} \ox \r_{A_3A_4}$ be a four-qubit diagonal state. Then $\r$ is 2-UDA if and only if one of the following conditions holds, 
(i) $\r_{A_1A_2}$ or $\r_{A_3A_4}$ is pure;
(ii) at least two of the 1-marginals of $\r$ are pure.
\end{proposition}
\begin{proof}
The "if part" is trivial. We show the "only if" part. Suppose $\r_{A_1A_2}$, $\r_{A_3A_4}$ are both mixed, and there is at most one 1-marginal that is pure. Without loss of generality, we assume that the 1-marginal $\r_{A_1}$ is pure merely. Then $\r$ is 2-UDA iff $\r_{A_2}\ox \r_{A_3A_4}$ is 2-UDA. From Proposition \ref{pro:rho2-udaDiagonald-dim}, $\r$ is not 2-UDA. 
Otherwise, none of the 1- and 2-marginals of $\r$ is pure. One can establish another state $\s$ which has a nonzero element, and is compatible with $\r$. This completes the proof. 
\end{proof}

We consider the state $\r=\r_{A_1}\ox \r_{A_2}\ox\r_{A_3A_4}$.  From Lemma \ref{le:lr+(1-l)s_not2uda} (i), we have the following fact.  If the four-qubit state $\r=\r_{A_1}\ox \r_{A_2}\ox\r_{A_3A_4}$ is 2-UDA, then $\r_{A_2}\ox \r_{A_3A_4}$ and $\r_{A_1}\ox \r_{A_3A_4}$ is 2-UDA. It is analyzed in Theorem \ref{th:rhoA1 ox A23 iff}. 
If one of $\r_{A_1}$ and $\r_{A_2}$ is pure, then $\r$ is 2-UDA iff $\r_{A_k}\ox \r_{A_3A_4}$ is 2-UDA for $k=1,2$. Suppose $\r_{A_k}$ is mixed in the subsequent analysis. 
From Theorem \ref{th:rhoA1 ox A23 iff}, $\r=\r_{A_1}\ox \r_{A_2}\ox\r_{A_3A_4}$ is 2-UDA implies that one of the three conditions holds,     
(a) $\r_{A_3}$ or $\r_{A_4}$ is pure, i.e. $\r=\ox_{k=1}^4\r_{A_k}$. Further, Proposition \ref{pro:diag4-qubit} implies that $\r_{A_3}$ and $\r_{A_4}$ are both pure since $\r_{A_1}$ and $\r_{A_2}$ are mixed.
(b) $\r_{A_3A_4}$ is pure.
(c) $\rank(\r_{A_3A_4})=2$ and  $\r_{A_3A_4}$ is LU equivalent  to $D\oplus 0$ with the positive diagonal elements of $D$. 
Note that (a) and (b) imply that $\r$ is 2-UDA. We derive that  (c) is not a sufficient condition for $\r$ to be 2-UDA, which is shown as follows.
\begin{lemma}
\label{le:fourqubit-not2uda}
Suppose $\r=\r_{A_1}\ox \r_{A_2}\ox\r_{A_3A_4}$  is a four-qubit state.  Here   $\rank(\r_{A_1})=\rank(\r_{A_2})=\rank(\r_{A_3A_4})=2$ and  $\r_{A_3A_4}$ is LU equivalent  to $D\oplus 0$ in \eqref{eq:rhoA23=D+0} with the positive diagonal elements of $D$.  Then $\r$ is not 2-UDA. 
\end{lemma}
\begin{proof}
Since $\r_{A_1}\ox \r_{A_2}$ is full-rank, there is $D=\diag\{0,\e_1,\e_2,0\}$ with $\e_1,\e_2$ small enough, such that $\xi:=\r_{A_1}\ox \r_{A_2}-D\geq 0$ by Lemma \ref{le:Y-ve x}. Then $\r=\xi\ox \r_{A_3A_4}+D\ox \r_{A_3A_4}$.
We show that 
\begin{eqnarray}
\label{eq:a=DoxRhoA34}
\a:= D\ox\r_{A_3A_4}
\end{eqnarray}
is not 2-UDA. 
From Lemma \ref{le:lr+(1-l)s_not2uda} (ii), it suffices to consider the range of $\r_{A_3A_4}$, up to LU equivalence,  is given as follows, for $zuv \neq 0$ and $nh\neq 0$,
\begin{eqnarray}
\label{eq:rangeA341}
\hspace{-0.5cm}&& \cR(\r_{A_3A_{4}})=\text{span}\{z\ket{00}, \; u\ket{01}+v\ket{10}\}, 
\\
\label{eq:rangeA342}
\hspace{-0.5cm}&& \cR(\r_{A_3A_{4}})=\text{span}\{m\ket{00}+n\ket{01}, \; g\ket{00}+h\ket{10}\}.
\end{eqnarray}
Suppose $\s$ is a four-qubit state compatible with $\a$. Then 
\begin{eqnarray}
\s=0_{4\times 4}\oplus H \oplus0_{4\times 4},
\end{eqnarray}
where $H=\bma \e_1\r_{A_3A_4} & B\\
B^\dg & \e_2\r_{A_3A_4}\ema$ for $B=B_1\oplus0$ and $\tr{B}=0$.
By some observations, we can always choose appropriate $B_1$ such that $\norm{B}$ is small enough and
\begin{eqnarray}
\label{eq:RBsubsetR34}
  &&\cR(B^\dg)\subseteq \cR(\r_{A_3A_4})
  \\
  \label{eq:RBBsubsetR34}
  &&
  \cR(BB^\dg)=\cR(B)\subseteq \cR(\r_{A_3A_4}),
\end{eqnarray}
where \eqref{eq:RBBsubsetR34} is implied by \eqref{eq:RBsubsetR34}. 
In detail, we consider the case for the range of $\r_{A_3 A_4}$ is given in
\eqref{eq:rangeA341}. We choose $$B_1=\bma -(y_{1}^{2}+y_{2}^{2}) & 0 & 0 \\
0 & y_{1}^{2} & y_{1}y_{2} \\
0 & y_{1}y_{2} & y_{2}^{2}\ema$$ for $y_i>0$ small enough and $\frac{y_1}{y_2}=\frac{u}{v}$.
Second, the range of $\r_{A_3 A_4}$ is given as in \eqref{eq:rangeA342}. For $mg\neq 0$, we choose $$B_1=\frac{1}{\mu}\bma x_{1}^{2}x_{4}^{2}-x_{2}^{2}x_{3}^{2} &
x_{1}x_{2}(x_{3}^{2}+x_{4}^{2}) &
-x_{3}\mu \\
x_{1}x_{2}(x_{3}^{2}+x_{4}^{2}) &
x_{2}^{2}(x_{3}^{2}+x_{4}^{2}) &
0 \\
-x_{3}\mu & 0 & -\,x_{4}\mu \ema$$ for $\mu=x_{4}(x_{1}^{2}+x_{2}^{2})$, $x_i>0$ small enough,  $\frac{x_1}{x_2}=\frac{m}{n}$ and $\frac{x_3}{x_4}=\frac{g}{h}$.

Since $\rank(\r_{A_3A_4})=2$, there is a unitary $V$ such that $H$ is congruent to $\bma \e_1\r_{A_3A_4} & BV\\
V^\dg B^\dg & I_2\oplus0_2\ema$. Using \eqref{eq:RBsubsetR34}, we obtain that $\cR(BV)\subset \cR(I_2\oplus 0_2)$, and thus $H$ is also congruent to
$$(\e_2\r_{A_3A_4}-B B^\dg)\oplus I_2 \oplus 0_2.$$ 
Using Lemma \ref{le:Y-ve x} and \eqref{eq:RBBsubsetR34}, for $x_i, y_i$ small enough, we have $(\e_2\r_{A_3A_4}- BB^\dg)\geq 0$, and hence $H\geq 0$. Then $\s\geq 0$. 
This shows that $\a$ is not 2-UDA. Using Lemma \ref{le:lr+(1-l)s_not2uda}, $\r$ is not 2-UDA. 
\end{proof}

By summarizing the analysis above Lemma \ref{le:fourqubit-not2uda},  we obtain that   $\r$ is 2-UDA iff condition (a) or (b) holds. This leads to the following fact.  
\begin{proposition}
\label{pro:4-qubit2UDA}
Let $\r=\r_{A_1}\ox \r_{A_2}\ox\r_{A_3A_4}$  be a four-qubit state. Then

(i) If at least one of $\r_{A_1}$ or $\r_{A_2}$ is pure, then $\r$ is 2-UDA iff the remaining three-qubit state $\r_{A_k}\ox \r_{A_3A_4}$ is 2-UDA given in Theorem \ref{th:rhoA1 ox A23 iff}.

(ii) If both $\r_{A_1}$ and $\r_{A_2}$ are mixed, then $\r$ is 2-UDA iff  $\r_{A_3A_4}$ is pure. 
\end{proposition}

We assume that $\r_{A_1A_2}$ is not necessarily product states, and show the fully characterization of 2-UDA states which is the tensor product of rank-two two-qubit states. From Lemma \ref{le:simplify2-qubit}, up to LU equivalence,  one can  assume that
\begin{eqnarray}
\label{eq:rho=rho12 ox rho 34}
\r&=&\r_{A_1A_2}\ox \r_{A_3A_4}
\\
&:=& 
\bma p_1 &b & c & 0\\
b^* &p_2&a &0 \\
c^*& a^* &p_3&0\\
0& 0& 0& 0\ema\ox 
\bma q_1 &y & z & 0\\
y^* &q_2&x &0 \\
z^*& x^* &q_3&0\\
0& 0& 0& 0\ema. 
\nonumber
\end{eqnarray} 
\begin{proposition}
\label{pro:rank4_rA12OxrA34}
Let $\r=\r_{A_1A_2}\ox \r_{A_3A_4}$ be a four-qubit state in \eqref{eq:rho=rho12 ox rho 34} with $\rank(\r_{A_1A_2})=\rank(\r_{A_3A_4})=2$. Then 

(i) If $p_2p_3=0$, then $\r=\r_{A_1}\ox \r_{A_2}\ox \r_{A_3A_4}$ has been investigated in Proposition \ref{pro:4-qubit2UDA}. 

(ii) If $p_2p_3>0$ and $p_1=0$ or $q_2q_3>0$ and $q_1=0$, $\r$ is not 2-UDA.

(iii) If $p_1p_2p_3>0$ and $q_1q_2q_3>0$, then $\r$ is 2-UDA. 
\end{proposition}
\begin{proof}
(i) By some observations, we obtain that $\r_{A_1A_2}=\r_{A_1}\ox \r_{A_2}$ iff $p_2p_3=0$. In this case,  $\r=\r_{A_1}\ox \r_{A_2}\ox \r_{A_3A_4}$ has been investigated in Proposition \ref{pro:4-qubit2UDA}. 
The same analysis applies to $\r_{A_3A_4}$.

(ii) We show the proof by two cases: (ii.a) $q_1=0$ and (ii.b) $q_1>0$.
(ii.a) If $p_1=0$, $\abs{a}<\sqrt{p_2p_3}$ and $q_1=0$, $\abs{x}<\sqrt{q_2q_3}$, then $\r$ is not 2-UDA. 
We establish a state $\s$  compatible with $\r$, where 
\begin{eqnarray}
\s=0_{5\times 5}\oplus H \oplus 0_{5\times 5}.
\end{eqnarray}
For $\abs{\e}>0$ small enough, $$H=\bma H_{11} &0& H_{12}\\ 0&0&0\\ H_{12}^{\dg} &0 & H_{22}\ema$$ 
for 
$H_{11}=\bma p_2q_2&p_2 x \\ (p_2x)^*&p_2q_3\ema$,
$H_{12}=\bma aq_2&ax+\e \\ (ax)^*&aq_3\ema$ and 
$H_{22}=\bma p_3q_2&p_3 x\\
(p_3x)^*&p_3q_3\ema$.
 One can verify that  $H\geq 0$. In fact, by discarding the zero rows and columns of $H$, the remaining matrix is given by $A=\bma p_2 &a\\ a^*& q_2\ema \ox 
\bma q_2 &x\\ x^*& q_3\ema>0$. Then the perturbation remains the positive semi-definiteness of $A$. Since $\e\neq 0$,    we have $\s\neq \r$. Therefore, $\r$ is not 2-UDA.  

(ii.b) $p_1=0$, $\abs{a}<\sqrt{p_2p_3}$ and $q_1q_2q_3>0$ (equivalently, $p_1p_2p_3>0$ and $q_1=0$, $\abs{x}<\sqrt{q_2q_3}$).   In this case, the state $\r$ is given by
\begin{eqnarray}
\r&=&\r_{A_1A_2}\ox \r_{A_3A_4}
\nonumber\\
&=&\bma 0&0&0&0\\
0&p_2 &a&0\\
0&a^*& p_3 &0\\
0&0&0&0 \ema \ox \bma q_1 &y & z & 0\\
y^* &q_2&x &0 \\
z^*& x^* &q_3&0\\
0& 0& 0& 0\ema. 
\nonumber
\end{eqnarray}
Since $\abs{a}<\sqrt{p_2p_3}$, there is a $D=\diag\{0,\e_1,\e_2,0\}$ with $\e_k>0$ small enough, such that
\begin{eqnarray}
\label{eq:rho=prho-D}
\r= (\r_{A_1A_2}-D)\ox \r_{A_3A_4}+ \a,
\end{eqnarray}
where $\a$ in \eqref{eq:a=DoxRhoA34} is not 2-UDA, and
$\r_{A_1A_2}-D\geq 0$. 
 Using Lemma \ref{le:lr+(1-l)s_not2uda} and \eqref{eq:rho=prho-D}, $\r$ is not 2-UDA.

(iii) 
We show that  $\r$ is 2-UDA by contradiction. Suppose $\r$ is not 2-UDA and $\s$ is a state compatible with $\rho$. 
Let $\chi:=\rho-\s$.
Then $\chi$ is a Hermitian operator
$\chi$ on $\cH_{A_1A_2A_3A_4}$ satisfying $\tr_{A_jA_k} \chi=0$. Obviously, $\rho$ is not 2-UDA iff $\chi\neq 0$ for some state $\s$. 
However, one can show that $\chi=0$ as follows. 
From Lemma \ref{le:general_properties} (v), we have  $\cR(\s)\subseteq \cR(\r)$, and hence
\begin{eqnarray}
\label{eq:RXsubsetRr}
\cR(\chi)\subseteq \cR(\r)
=\text{span}\{\ket{a_i}:i=1,2,3,4\}.
\end{eqnarray}
Let $A=(\ket{a_1}, \ket{a_2}, \ket{a_3}, \ket{a_4})$    be a matrix with rank equal to four.
Since $\chi$ is Hermitian and $\cR(\chi)\subseteq \cR(\r)$, we claim that without loss of generality,  
\begin{eqnarray}
\label{eq:chi=ASAdg}
\chi=ASA^\dg,
\end{eqnarray}
where $S=(s_{kl})\in \bbC^{4\times 4}$ is a  Hermitian matrix for $k,l=1,2,3,4$.  From $\chi=\chi^\dg$, we have $S=:(A^\dg A)^{-1} A^\dg \chi A (A^\dg A)^{-1}$ is Hermitian. One can verify that $ASA^\dg=\chi$ as follows. First $P_A=A(A^\dg A)^{-1} A^\dg$ is a projector on $\cR(A)=\cR(\r)$ since $P_Ax=x$, $\forall x\in \cR(A)$ and $P_A$ is self-adjoint and 
idempotent. From $\cR(\chi)\subset \cR(\r)$, we have $P_A\chi =\chi$, and $\chi P_A=\chi$ by Hermitian $\chi$. This implies that $ASA^\dg=P_A\chi P_A=\chi$. This proves the claim. 
Note that $p_1p_2p_3>0$ and $q_1q_2q_3>0$. Up to LU equivalence, the range of $\r_{A_iA_{i+1}}$, for $i=1,3$, is one of the following 
\begin{eqnarray}
\label{eq:RrhoA341}
&& \cR(\r_{A_iA_{i+1}})=\text{span}\{m_i\ket{00}+n_i\ket{01}, 
\\
&& \hspace{3.2cm}g_i\ket{00}+h_i\ket{10}\},\nonumber\\
\label{eq:RrhoA342}
&& \cR(\r_{A_iA_{i+1}})=\text{span}\{z_i\ket{00}, \; u_i\ket{01}+v_i\ket{10}\}, 
\end{eqnarray}
where $n_ih_i\neq 0$ and $z_iu_iv_i \neq 0$ since $\r_{A_iA_{i+1}}$ has two diagonal elements if they are zero, and $z_i\neq 0$ by $\rank(\r_{A_iA_{i+1}})=2$.

The proof of $\chi=0$ above
consists of three parts (iii.a)-(iii.c), depending on the range of $\r_{A_1A_2}$ and $\r_{A_3A_4}$. 

(iii.a)  $\cR(\r_{A_1A_2})$ and $\cR(\r_{A_3A_4})$ follow  \eqref{eq:RrhoA341}. Then $\ket{a_i}$'s in \eqref{eq:RXsubsetRr} are 
\begin{eqnarray}
\label{eq:keta1}
&& \hspace{-0.5cm}
\ket{a_1}=(m_1\ket{00}+n_1\ket{01})\ox (m_3\ket{00}+n_3\ket{01}),
\\
&& \hspace{-0.5cm}
\ket{a_2}=(m_1\ket{00}+n_1\ket{01})\ox (g_3\ket{00}+h_3\ket{10}),
\\
&& \hspace{-0.5cm}
\ket{a_3}=(g_1\ket{00}+h_1\ket{10})\ox (m_3\ket{00}+n_3\ket{01}),
\\
\label{eq:keta4}
&& \hspace{-0.5cm}
\ket{a_4}=(g_1\ket{00}+h_1\ket{10})\ox (g_3\ket{00}+h_3\ket{10}).
\end{eqnarray}
From $\rank(\r_{A_1A_2})=\rank(\r_{A_3A_4})=2$, at most one element of the two sets $\{m_i,g_i\}$ is zero, respectively.

We show that  $\tr_{A_jA_k}\chi =0$  implies $S=0$ as follows. Let $\chi=(H_{ij})$, where $H_{i,j}\in \bbC^{2\times 2}$ and $i,j=1,2,...,8$. By \eqref{eq:chi=ASAdg}, we have $H_{j,k}=0$ if $j,k=7,8$. 
From $\chi_{A_1A_2}=\chi_{A_1A_3}=\chi_{A_2A_3}=0$ and $H_{77}=H_{88}=0$, we have $\tr{H_{jj}}=0$ for $j=1,2,...,8$. In particular, we have $s_{22}=s_{44}=0$ by $\tr(H_{44})=\abs{h_3}^2 \abs{n_1}^2 s_{22}=0$ and $\tr(H_{66})=\abs{h_1}^2 \abs{h_3}^2s_{44}=0$. Next we show that $s_{34}=0$. From $s_{22}=s_{44}=0$ and $\chi_{A_1A_4}=0$, we have $H_{55}=0$, and thus $\abs{h_1}^2g_3^*n_3s_{34}=0$. By $\chi_{A_1A_3}=0$, we have $\tr(H_{56})=\abs{h_1}^2h_3^*m_3 s_{34}=0$.
Since $m_3$ and $g_3$ cannot be zero simultaneously,
we have $s_{34}=0$. Next, we have $s_{24}=0$ since $\chi_{A_1A_3}=0$ and $\chi_{A_2A_3}=0$ derive that $\tr(H_{26})=h_1^*\abs{h_3}^2 m_1s_{24}=0$ and  
$\tr(H_{24})=\abs{h_3}^2 n_1^* g_1 s_{24}=0$. Note that $m_1$ and $g_1$ cannot be zero at the same time. Hence $s_{24}=0$.
Besides, $\chi_{A_1A_4}=\chi_{A_2A_4}=0$ derives that $H_{33}=H_{55}=0$, and thus $\abs{n_1}^2\abs{n_3}^2s_{11}=\abs{h_1}^2\abs{n_3}^2s_{33}=0$. We have $s_{11}=s_{33}=0$. 
Using $\chi_{A_1A_4}=\chi_{A_2A_4}=0$, we have $H_{15}=H_{13}=0$, and thus $\abs{n_3}^2h_1^*m_1s_{13}=\abs{n_3}^2n_1g_1^*s_{13}=0$. We obtain that $s_{13}=0$.
Next we show that $s_{12}=0$. By $\chi_{A_2A_4}=\chi_{A_2A_3}=0$, we have $H_{33}=0$ and $\tr(H_{34})=0$, which implies $\abs{n_1}^2n_3g_3^*s_{12}=\abs{n_1}^2h_3^*m_3 s_{12}=0$. As a complicated case, we show that $s_{14}=0$. From $\chi_{A_2A_3}=0$, we have $\tr(H_{14})=h_3^*n_1^*g_1m_3s_{14}=0$, $\tr(H_{23})=h_3^*n_1 g_1^*m_3 s_{14}=0$ and $H_{13}=0$, which derives $n_1n_3g_1^*g_3^* s_{14}=0$. Using $\chi_{A_1A_4}=0$, we have $H_{15}=0$ and thus $n_3h_1^*m_1g_3^* s_{14}=0$. By $\chi_{A_1A_3}=0$, we have $\tr{(H_{16})}=h_1^*h_3^*m_1m_3 s_{14}=0$. Note that at most one element of the two sets $\{m_i,g_i\}$ is zero, respectively. Hence $s_{14}=0$. Finally, we show that  $s_{23}=0$. From $\chi_{A_2A_4}=\chi_{A_1A_4}=0$, we have $H_{13}=H_{15}=0$, and thus $n_1n_3^*g_1^*g_3 s_{23}=h_1^*n_3^*m_1g_3 s_{23}=0$. 
By $\chi_{A_1A_3}=\chi_{A_2A_3}=0$, we have $\tr(H_{25})=h_1^*h_3m_1m_3^* s_{23}=0$ and $\tr(H_{14})=h_3n_1g_1^*m_3^* s_{23}=0$. We have $s_{23}=0$.
To conclude, we show that the Hermitian operator $S=0$, and thus $\chi=\r-\s=0$ for any $\s$. It is a contradiction.
Therefore, $\r$ is 2-UDA.

(iii.b)  $\cR(\r_{A_1A_2})$ and $\cR(\r_{A_3A_4})$ follow \eqref{eq:RrhoA342}. By the same analysis as (iii.a), one can verify that $S=0$ and thus $\chi=0$ for any $\s$, which is shown in TABLE \ref{tab:S_element}. The state $\r$ is 2-UDA in this case.  
 
(iii.c) $\cR(\r_{A_1A_2})$ follows \eqref{eq:RrhoA342} and $\cR(\r_{A_3A_4})$ follows \eqref{eq:RrhoA341}, up to system permutation of $A_1A_2$ and $A_3A_4$. We obtain that $S=0$ in TABLE \ref{tab:S_element}, and  $\chi=ASA^\dg=0$. Therefore, $\r$ is 2-UDA. This completes the proof.
\end{proof}

\begin{table*}
    \centering
    \begin{tabular}{|c|c|}
\hline
 $\cR(\r_{A_1A_2})$ and $\cR(\r_{A_3A_4})$ follow  \eqref{eq:RrhoA342}.   & $\cR(\r_{A_1A_2})$ follows \eqref{eq:RrhoA342} and $\cR(\r_{A_3A_4})$ follows \eqref{eq:RrhoA341} \\ 
\hline
$\tr(H_{22})=\abs{v_3}^2\abs{z_1}^2s_{22}=0 \Rightarrow s_{22}=0$        & 
$\tr(H_{22})=\abs{h_3}^2\abs{z_1}^2s_{22}=0 \Rightarrow s_{22}=0$ \\
\hline
$\tr(H_{44})=\abs{v_3}^2\abs{u_1}^2s_{44}=0 \Rightarrow s_{44}=0$          & 
$\tr(H_{66})=\abs{v_1}^2\abs{h_1}^2s_{44}=0 \Rightarrow s_{44}=0$ \\
\hline        
$\tr(H_{14})=u_1^*v_3^*z_1z_3s_{14}=0 \Rightarrow s_{14}=0$          & 
$\tr(H_{26})=\abs{h_3}^2v_1^*z_1s_{24}=0 \Rightarrow s_{24}=0$  \\
         \hline
$\tr(H_{26})=\abs{v_3}^2v_1^*z_1s_{24}=0 \Rightarrow s_{24}=0$         &
$H_{15}=0\Rightarrow \abs{n_3}^2 z_1 v_1^*s_{13}=0 \Rightarrow s_{13}=0$  \\
         \hline
$\tr(H_{25})=v_1^*v_3z_1z_3^*s_{23}=0 \Rightarrow s_{23}=0$         & 
$\left. \begin{array}{l}
\tr(H_{14})=u_1^*h_3^*z_1m_3s_{14}=0 \\
H_{13}=0 \Rightarrow u_1^*z_1n_3g_3^*s_{14}=0
\end{array} \right\}
\Rightarrow s_{14}=0$
\\
         \hline
$\tr(H_{35})=u_1v_1^*\abs{z_3}^2s_{33}=0 \Rightarrow s_{33}=0$         &
$\left. \begin{array}{l}
\tr(H_{25})=v_1^*h_3z_1m_3^*s_{23}=0 \\
H_{15}=0 \Rightarrow n_3^*v_1^*z_1g_3s_{23}=0
\end{array} \right\}
\Rightarrow s_{23}=0$
\\
       \hline
 $H_{11}=0 \Rightarrow \abs{z_1}^2 \abs{z_3}^2 s_{11}=0 \Rightarrow s_{11}=0$          &
 $H_{55}=0\Rightarrow
\abs{n_3}^2\abs{v_1}^2s_{33}=0\Rightarrow s_{33}=0$\\
         \hline
$H_{11}=0 \Rightarrow \abs{z_1}^2 z_3u_3^* s_{12}=0 \Rightarrow s_{12}=0$         
& 
$H_{11}=0\Rightarrow
\abs{n_3}^2\abs{z_1}^2s_{11}=0\Rightarrow s_{11}=0$\\
         \hline
$H_{13}=0 \Rightarrow \abs{z_3}^2 z_1 u_1^* s_{13}=0 \Rightarrow s_{13}=0$           
&
$\left. \begin{array}{l}
\tr(H_{34})=\abs{u_1}^2n_3^* g_3s_{34}=0 \\
H_{33}=0 \Rightarrow \abs{u_1}^2 h_3^*m_3s_{34}=0
\end{array} \right\}
\Rightarrow s_{34}=0$
\\
         \hline         
$H_{55}=0 \Rightarrow \abs{v_1}^2 z_3u_3^* s_{34}=0 \Rightarrow s_{34}=0$ 
& 
$\left. \begin{array}{l}
H_{11}=0 \Rightarrow  \abs{z_1}^2 n_3g_3^*s_{12}=0
 \\
\tr(H_{12})=\abs{z_1}^2h_3^*m_3 s_{12}=0
\end{array} \right\}
\Rightarrow s_{12}=0$
\\
         \hline
    \end{tabular}
    \caption{The elements $s_{kl}$ of $S$ given in \eqref{eq:chi=ASAdg}, under the condition $\tr_{A_jA_k}\chi=0$, where $\chi=(H_{ij})$ and $H_{i,j}\in \bbC^{2\times 2}$ for $i,j=1,2,...,8$}
\label{tab:S_element}
\end{table*}

We show that the tensor product of two mixed states is not 2-UDA whenever one of them has rank not less than three and the other is not pure.
\begin{proposition}
\label{pro:4qubit-highrank}
 If one of $\rank(\r_{A_1A_2})$ and $\rank(\r_{A_3A_4})$ has rank not less than three, and the other is not pure, then $\r=\r_{A_1A_2}\ox \r_{A_3A_4}$ is not 2-UDA. 
\end{proposition}
\begin{proof}
Suppose $\rank(\r_{A_1A_2})=2$ and $\rank(\r_{A_3A_4})=3$. Then $\r$ is given by
\begin{eqnarray}
\label{eq:rho=rho2 ox rho 34}
\r&=&\r_{A_1A_2}\ox \r_{A_3A_4}
\\
&:=& 
\bma p_1 &b & c & 0\\
b^* &p_2&a &0 \\
c^*& a^* &p_3&0\\
0& 0& 0& 0\ema\ox 
\r_{A_3A_4}.
\nonumber
\end{eqnarray}
The UDA property of $\r$ can be analyzed by the following cases.

(i) If $p_2p_3=0$, then $\r=\r_{A_1}\ox \r_{A_2}\ox \r_{A_3A_4}$ and $\r$ is not 2-UDA since $\r_{A_2}\ox \r_{A_3A_4}$ is not 2-UDA.

(ii) $p_1=0$ and $p_2p_3>0$; 
Let 
\begin{eqnarray}
\label{eq:diagonalD}
D=\diag\{0,\e_1,\e_2,0\}
\end{eqnarray}
 with $\e_i>0$ small enough.
We show that $\r$ is not 2-UDA since 
$\r_{A_1A_2}=(\r_{A_1A_2}-D)+D:=\a_1+\a_2$, for $\a_1\geq 0$. By similar analysis as \eqref{eq:a=DoxRhoA34},  we obtain that $\a_2\ox \r_{A_3A_4}$ is not 2-UDA. In detail, a state $\s$ compatible with $\r$ is given by $\s=0_2\oplus \bma \e_1 \r_{A_3A_4} & B \\ B^\dg &\e_2 \r_{A_3A_4}\ema \oplus 0_2$ with $\tr B=0$. We can choose $B=\diag\{0,x,-x,0\}$ for $x>0$ small enough, such that $\cR(B)\subset \cR(\r_{A_3A_4})$. Then $\s\neq \r$ is compatible with $\rho$.   From Lemma \ref{le:lr+(1-l)s_not2uda}, $\r$ is not 2-UDA. 

(iii) $p_1p_2p_3>0$. 
Since $\r_{A_3A_4}$ is rank-three, we claim that up to LU equivalence, $\cR(D)\subseteq \cR(\r_{A_3A_4})$, where $D$ is given in \eqref{eq:diagonalD}. In fact, if $\r_{A_3A_4}$ is LU equivalent to a state with a zero diagonal element, we have $\r_{A_3A_4}=N\oplus0$ without loss of generality, where $N$ is full rank. The claim holds. 
Otherwise, $\r_{A_3A_4}$
 is not LU equivalent to the state with a zero diagonal element. The range of $\r_{A_3A_4}$ is 
$
\cR(\r_{A_3A_4})=\text{span}\{\ket{\ps_1}:=\ket{01}, \ket{\ps_2}:=\ket{10}, \ket{\ps}:=\sin \t\ket{00}-\cos\t\ket{11} \}$. We have proven the claim. 
Then we have $\r=\r_{A_1A_2}\ox (\r_{A_3A_4}-D)+\r_{A_1A_2}\ox D:= \b_1+\b_2$. We showed that $\b$ is not 2-UDA in the analysis below \eqref{eq:a=DoxRhoA34}. Using Lemme \ref{le:lr+(1-l)s_not2uda}, $\r$ is not 2-UDA.  

Next we show that $\r$ is not 2-UDA if $\rank(\r_{A_1A_2})=2$ and $\rank(\r_{A_3A_4})=4$. 
The spectral decomposition of $\r_{A_3A_4}=(\sum_{i=1}^3 \l_i\proj{a_i})+\l_4\proj{a_4}:=p\a+(1-p)\b$, where $p=\sum_{i=1}^3\l_i$ and $\a=\frac{1}{p}\sum_{i=1}^3 \l_i\proj{a_j}$ is rank-three. Then $\r=p\r_{A_1A_2}\ox \a+(1-p)\r_{A_1A_2}\ox \b$, where the first term $\r_{A_1A_2}\ox \a$ is not 2-UDA by the preceding analysis.  Using Lemma \ref{le:lr+(1-l)s_not2uda}, $\r$ is not 2-UDA. Similarly, one can show that $\r$ is not 2-UDA if $\rank(\r_{A_1A_2})\geq 3$ and $\rank(\r_{A_3A_4})\geq 3$. The UDA property is invariant up to system permutations between systems $A_1A_2$ and $A_3A_4$. This completes the proof.
\end{proof}

\section{Auxiliary results for $n$-qubit mixed UDA states}
\label{app:n-qubit}

The following fact excludes a set of states that is 2-UDA by establishing  different states that are compatible with them. 
\begin{lemma}
\label{le:rA1oxrA2345}
Suppose $\r_{A_1}$ is a mixed one-qubit state, the two-qubit states $\r_{A_2A_3}$ and $\r_{A_4A_5}$ satisfy the condition in Theorem \ref{th:RA12oxRA34} (iv). Then 
the five-qubit state   
$\r=\r_{A_1}\ox \r_{A_2A_3} \ox \r_{A_4A_5}$ is not 2-UDA. 
\end{lemma} 
\begin{proof}
 Suppose $\s$ is a state 2-compatible with $\r$. By Lemma \ref{le:lr+(1-l)s_not2uda} (v),   $\s$ further satisfies 
 \begin{eqnarray}
 \label{eq:sA2345}
&&\s_{A_2A_3A_4A_5}=
\r_{A_2A_3}\ox \r_{A_4A_5},
\\
&& 
 \label{eq:sA123}
\s_{A_1A_2A_3}=\r_{A_1}\ox \r_{A_2A_3}, 
\\
&& 
 \label{eq:sA145}
\s_{A_1A_4A_5}=\r_{A_1}\ox \r_{A_4A_5}.
 \end{eqnarray}
 Obviously, these conditions imply that $\s$ is 2-compatible with $\r$. 
From \eqref{eq:sA2345}, any $\s$ can be given by
\begin{eqnarray}
\s_{A_1...A_5}=\bma X_{A_2...A_5} & Y_{A_2...A_5}\\
Y_{A_2...A_5}^\dg & \r_{A_2A_3}\ox \r_{A_4A_5}-X_{A_2...A_5} \ema,
\end{eqnarray}
where by \eqref{eq:sA123} and \eqref{eq:sA145},
we have $
 0\leq X\leq \r_{A_2A_3}\ox \r_{A_4A_5}$, 
$ X_{A_2A_3}=h_1\r_{A_2A_3}, \; 
X_{A_4A_5}=h_1\r_{A_4A_5}$,
and 
$Y_{A_2A_3}=Y_{A_4A_5}=0$.
Then $\r$ is 2-UDA iff $\s=\r_{A_1}\ox\r_{A_2A_3}\ox \r_{A_4A_5}$ iff $X=h_1\r_{A_2A_3}\ox \r_{A_4A_5}$ and $Y=0$. 
We establish a different $\s$ that is compatible with $\r$, such that $X=h_1\r_{A_2A_3}\ox \r_{A_4A_5}$ but $Y\neq 0$. Then 
\begin{eqnarray}
\label{eq:sigma_Y}
\s=\bma h_1\r_{A_2A_3}\ox \r_{A_4A_5} & Y\\
Y^\dg &  h_2\r_{A_2A_3}\ox \r_{A_4A_5}\ema.  
\end{eqnarray}
 In order to make $\s\geq 0$, we choose nonzero Hermitian $Y$ satisfying $\cR(Y)\subseteq \cR(\r_{A_2A_3}\ox \r_{A_4A_5})$, $Y_{A_2A_3}=Y_{A_4A_5}=0$ and $\norm{Y}$ small enough. By the same analysis as Lemma \ref{le:fourqubit-not2uda}, $\s$ is congruent to $I_4\oplus 0_{12}\oplus (h_2\r_{A_2A_3}\ox \r_{A_4A_5}-Y^\dg Y)$, which is positive semidefinite. We show the expression of $Y$.
Since $\cR(Y)\subseteq  \cR(\r_{A_2A_3}\ox \r_{A_4A_5})$, we have $Y=ASA^\dg$, where $S=(s_{jk})$ for $j,k=1,2,3,4$ is Hermitian, and $A=(\ket {a_1}, \ket {a_2}, \ket {a_3}, \ket {a_4})$. Here $\ket{a_i}$'s are the base of the vector space $\cR(\r_{A_2A_3}\ox \r_{A_4A_5})$. 
Depending on the range of $\cR(\r_{A_2A_3}\ox \r_{A_4A_5})$, this is derived by three cases.
First, $\cR(\r_{A_2A_3})$ and  $\cR(\r_{A_4A_5})$ both follow \eqref{eq:RrhoA341}.
Then $\ket{a_i}$'s are given in \eqref{eq:keta1}-\eqref{eq:keta4}, and $S$ is diagonal with $s_{11}=x$, 
$s_{22}=-\frac{\abs{m_3}^2+\abs{n_3}^2}{\abs{g_3}^2+\abs{h_3}^2}x$, 
$s_{33}=-\frac{\abs{m_1}^2+\abs{n_1}^2}{\abs{g_1}^2+\abs{h_1}^2}x$, 
$s_{44}=-\frac{(\abs{m_1}^2+\abs{n_1}^2)(\abs{m_3}^2+\abs{n_3}^2)}{(\abs{g_1}^2+\abs{h_1^2})(\abs{g_3}^2+\abs{h_3^2})}x$, for $x>0$ small enough.  
Second, $\cR(\r_{A_2A_3})$ and  $\cR(\r_{A_4A_5})$ both follow \eqref{eq:RrhoA342}. Then $\ket{a_1}=z_1z_3\ket{00}$, $\ket{a_2}=z_1\ket{00}\ox (u_3\ket{01}+v_3\ket{10})$, 
$\ket{a_3}=(u_1\ket{01}+v_1\ket{10})\ox z_3\ket{00}$,
$\ket{a_4}=(u_1\ket{01}+v_1\ket{10})\ox (u_3\ket{01}+v_3\ket{10})$. We choose $s_{12}=(\abs{u_1}^2+\abs{v_1}^2)x$, $s_{34}=-\abs{z_1}^2 x$, and other elements of $S$ are zero, for $x>0$ small enough. 
Third, up to system permutation of $A_2A_3$ and $A_4A_5$,
$\cR(\r_{A_2A_3})$ and $\cR(\r_{A_4A_5})$ follow \eqref{eq:RrhoA342} and \eqref{eq:RrhoA341}, respectively. Then 
$\ket{a_1}=z_1\ket{00}\ox (m_3\ket{00}+n_3\ket{01})$,
$\ket{a_2}=z_1\ket{00}\ox (g_3\ket{00}+h_3\ket{10})$,
$\ket{a_3}=(u_1\ket{01}+v_1\ket{10})\ox (m_3\ket{00}+n_3\ket{01})$,
and 
$\ket{a_4}=(u_1\ket{01}+v_1\ket{10})\ox (g_3\ket{00}+h_3\ket{10})$. 
We choose
$s_{13}=-(\abs{g_3}^2+\abs{h_3}^2) x$, 
$s_{24}=(\abs{m_3}^2+\abs{n_3}^2) x$ and other elements of $S$ are zero, for $x>0$ small enough. 
To conclude, we can always find a nonzero $Y$ in \eqref{eq:sigma_Y} such that $\s$ is 2-compatible with $\r$. This completes the proof. 
\end{proof}

We present the characterization of some $n$-qubit $k$-UDA states. 
We consider the fully product $k$-UDA state $\ox_{i=1}^n\r_{A_i}$. Next we analyze two general cases $\r=\ox_{i=1}^{n-2}\r_{A_i}\ox \r_{A_{n-1}A_n}$ and $\r=\ox_{i=1}^{n-4}\r_{A_i}\ox \r_{A_{n-3}A_{n-2}}\ox \r_{A_{n-1}A_n}$ for 2-UDA states.
Its proof is given as follows.  

\begin{proof}
(i) The "if part" is trivial by Lemma \ref{le:pure sep}. We show the "only if" part as follows. Suppose there are only $n-k-1$ of the $\r_{A_i}$'s that are pure. From Lemma \ref{le:lr+(1-l)s_not2uda} (iv), without loss of generality, we assume that $\r_{A_j}=\proj{\ps^{(j)}}$ for $j=1,2,...,n-k-1$ are pure.
For any state $\s$ $k$-compatible with $\r$, we have 
$\s=\ox_{i=1}^{n-k-1} \proj{\ps^{(j)}}_{A_j} \ox \s_{A_{n-k}, ..., A_n}$. 
It suffices to determine whether the $(k+1)$-partite state  $\s_{A_{n-k}, ..., A_n}$ is $\ox_{i=n-k}^n\r_{A_i}$ for any $\s$ and mixed state $\r_{A_i}$. 
We show that a $(k+1)$-partite state $\l=\ox_{i=1}^{k+1}\r_{A_i}$ is not $k$-UDA by induction. The case for $k=1$ holds by Proposition \ref{pro:bipartite1-UDA}. Suppose the fact holds for $k$-partite state.   
We show that it is also hold for a $k+1$-partite state $\l=\ox_{i=1}^{k+1}\r_{A_i}$. Since $\l_{A_1...A_k}=\ox_{i=1}^{k}\r_{A_i}$ is not $(k-1)$-UDA, there is a different state $\s'$ that is $(k-1)$-compatible with $\l_{A_1...A_k}$. Then $\s=\s'\ox \r_{A_{k+1}}$ is $k$-compatible with $\l$, and $\s\neq \l$. This proves the claim. 
Hence, $\s_{A_{n-k}, ..., A_n}$ is not necessarily $\ox_{i=n-k}^n\r_{A_i}$.

(ii) From Lemma \ref{le:lr+(1-l)s_not2uda} (i), $\r$ is 2-UDA only if $\r_{A_{n-2}}\ox \r_{A_{n-1}A_n}$ is 2-UDA. Using Theorem \ref{th:rhoA1 ox A23 iff}, this derives four cases, (a) $\r_{A_{n-1}A_n}$ is 2-UDA. Then $\r$ is 2-UDA iff $\ox_{i=1}^{n-2}\r_{A_{i}}$ is 2-UDA;  
(b) $\r_{A_{n-1}}$ or $\r_{A_n}$ is pure. By Lemma \ref{le:pure sep}, $\r=\ox_{i=1}^n\r_{A_i}$, which is considered in (i); 
(c) $\r_{A_{n-2}}$ is pure. If $\ox_{i=1}^{n-2}\r_{A_i}$ is pure, then $\r$ is 2-UDA for any $\r_{A_{n-1}}\r_{A_n}$; 
(d) if less than $n-2$ states of $\r_{A_i}$ are pure for $i=1,2,...,n-2$, then $\r$ is 2-UDA only when $\r_{A_{n-1}A_n}$ is rank-two and LU equivalent to $D\oplus 0$. Otherwise, there is a mixed $\r_{A_k}$ for $k=1,2,...,n-2$ such that $\r_{A_k}\ox \r_{A_{n-1}A_{n}}$ is not 2-UDA.  We claim that $\r=\ox_{i=1}^{n-2}\r_{A_i}\ox (D\oplus 0)$ is 2-UDA iff at least $n-3$ states of $\r_{A_i}$ are pure. It suffices to show the "only if" part. Suppose only $n-4$ states of $\r_{A_i}$ are pure, and, without loss of generality, they are $\r_{A_i}=\proj{\ps^{(i)}}_{A_i}$, for $i=1,2,...,n-4$.  Then any state $\s$ compatible with $\r$ is $\s=\ox_{i=1}^{n-4}\proj{\ps^{(i)}}_{A_i}\ox \s_{A_{n-3}...A_n}$. Then $\r$ is 2-UDA iff $\s_{A_{n-3}...A_n}=(\r_{A_{n-3}}\ox \r_{A_{n-2}})\ox (D\oplus 0)$, where $\r_{A_{n-3}}$ and $\r_{A_{n-2}}$ are mixed. By Theorem \ref{th:RA12oxRA34}, we obtain that $(\r_{A_{n-3}}\ox \r_{A_{n-2}})\ox (D\oplus 0)$ is not 2-UDA.  Hence $\r$ is not 2-UDA. 

(iii) $\r$ is 2-UDA only if $\r_{A_{n-3}A_{n-2}}\ox \r_{A_{n-1}A_n}$ is 2-UDA. This derives three cases using Theorem \ref{th:RA12oxRA34}. (a) $\r_{A_{n-3}A_{n-2}}$ or  $\r_{A_{n-1}A_n}$ is pure, then $\r$ is 2-UDA iff the rest part of the state is 2-UDA by (ii); (b) at least one of $\r_{A_{j}}$ is pure, the form of $\r$ reduces to (ii); (c) $\r_{A_{n-3}A_{n-2}}$ and $\r_{A_{n-1}A_{n}}$ are rank-two, 
$\r_{A_{n-3}A_{n-2}}\ox \r_{A_{n-1}A_n}$ is LU equivalent to  $(H\oplus0) \ox (D\oplus0)$. By Theorem \ref{th:5-qubit}, $\r_{A_j}\ox \r_{A_{n-3}A_{n-2}}\ox \r_{A_{n-1}A_n}$ is not 2-UDA if $\r_{A_j}$ is rank-two, for $j=1,2,...,n-4$. Hence, $\ox_{j=1}^{n-4}\r_{A_j}$ is 2-UDA, which implies that $\r$ is 2-UDA iff $\r_{A_{n-3}A_{n-2}}\ox \r_{A_{n-1}A_n}$ is 2-UDA.    
This completes the proof. 
\end{proof}

We exclude a set of high-rank $n$-partite mixed states to be the $k$-UDA in Proposition \ref{pro:rankn-1}. The proof is given as follows.

\begin{proof}
(i) For the full-rank $\r$, we can choose a nonzero Hermitian operator $\chi$ satisfying $\tr_{A_{i_1}...A_{i_k}}(\chi)=0$, $\forall i_j\in \{1,2,...,n\}$. We further choose $\s=\r+\e \chi$, where $\e$ is small enough to guarantee that $\s>0$.  
Thus $\r$ is not $k$-UDA and compatible with another full-rank state $\s$. 

Suppose $\rank(\r)= d_1d_2...d_n-1$.  Firstly, we consider the $n$-qubit state $\r$ for $\rank\r=2^n-1$. We prove that $\r$ is not 2-UDA for $n=3$. We claim that the range of $\r$ has a pure state $\ket{\g}$ LU equivalent to the quasi-GHZ state $a\ket{000}+b\ket{111}$ with $ab\ne0$. Up to LU equivalence, the kernel of a seven-dimensional space $V\subseteq \bbC^8$ satisfies that $V^\perp=\text{span}\{a_0\ket{000}+a_1\ket{001}+a_2\ket{010}+e^{i\t}a_3\ket{011}+a_4\ket{111}\}$ \cite{acin2000generalize}. The state $\ket{\b}=a_2^*\ket{001}+\frac{a_1^*a_2}{x^*}\ket{110}-a_1^*\ket{010}+x\ket{101}$ is LU equivalent to the quasi-GHZ state $a\ket{000}+b\ket{111}$ with $ab\neq 0$ and $\ket{\b}\perp V^\perp$. Therefore, the non-2-UDA state $\ket{\b}\in V$.  
It remains to prove the claim for $n$-qubit states $\r$ of rank $2^n-1$ and $n>3$. It suffices to show that the range of $\r$ has a pure state LU to the state $\ket{\g}\ket{0}^{\otimes (n-3)}$. Let $\ker\r$ be spanned by the pure state $\ket{\ps}=\sum_{(j_1,...,j_{n-3})\ne(0,...,0)}\ket{\a_{j_1...j_{n-3}}}\ket{j_1...j_{n-3}}+\ket{\b}\ket{0}^{\otimes n-3}$. Using the proof for $n=3$, we can assume that $\ket{\b}$ is orthogonal to $\ket{\g}$. Hence $\ket{\ps}$ is orthogonal to $\ket{\g}\ket{0}^{\otimes (n-3)}$. 

Next we consider the case for $\rank(\r)=d_1d_2...d_n-1$. Let $\ket{\ps}$ be a state supported on $\ox_{k=1}^n \bbC^{d_k}$.  Then $\ket{\ps}=(P+Q)\ket{\ps}$, where $Q$ is the projector onto the $n$-qubit space $H_Q$, and $P+Q=I_{\prod_j d_j}$. 
Let $S=\text{span}\{Q\ket{\ps} \}^\perp$ regarding to $H_Q$. We have $\dim S=2^n-1$, and $\forall \ket{x}\in S$, $\braket{x}{\ps}=0$, i.e. $S=\text{span}\{\ket{\ps} \}^\perp$.  By taking $\ket{\ps}$ as the vector orthogonal to $\cR(\r)$, we have $S\subseteq\cR(\r)$. By the preceding analysis, $S$ contains a pure state  LU equivalent to the quasi-GHZ state $a\ket{000}+b\ket{111}$ with $ab\neq 0$. Hence $\r$ is not $k$-UDA by Lemma \ref{le:lr+(1-l)s_not2uda} (i).
We have proven the claim.

(ii) Similar to the proof of Proposition \ref{pro:3qubit_general}, the range of $\r_{A_1...A_n}$ is the same as that of $\r_{A_1}\ox \r_{A_2A_3...A_n}$. By the result in (i), $\r_{A_2A_3...A_n}$ is not $k$-UDA for $1\leq k\leq n-2$. Using Lemma \ref{le:lr+(1-l)s_not2uda} (i) and (iii), $\r_{A_1...A_n}$ is not $k$-UDA. 
\end{proof}

\section*{Acknowledgment}
 This work is supported by   the NNSF of China (Grant No. 12471427) and the NUS research grant A-8003570-00-00. 

\bibliographystyle{IEEEtran}
\bibliography{udp}

\begin{IEEEbiographynophoto}{Xinyu Qiu}
received the B.S. degree in mathematics and applied mathematics from Shandong Normal University, Jinan, China, in 2019. She is currently pursuing the Ph.D. degree in applied mathematics with the School of Mathematical Sciences, Beihang University, Beijing, China, and is a visiting Ph.D. student with the Department of Mathematics, National University of Singapore, Singapore. Her research interests include quantum information theory, quantum entanglement, and quantum computation.
\end{IEEEbiographynophoto}

\begin{IEEEbiographynophoto}{Lin Chen}
received the B.S. degree in physics from Zhejiang University, Hangzhou, China, in 2003, and the Ph.D. degree in theoretical physics from Zhejiang University, Hangzhou, China, in 2008. He is currently an Associate Professor with the School of Mathematical Sciences, Beihang University, Beijing, China. Before joining Beihang University, he held research positions at several international institutions, including the Centre for Quantum Technologies, National University of Singapore, the Institute for Quantum Computing, University of Waterloo, Canada, and the Singapore University of Technology and Design. His research interests include quantum information theory, quantum entanglement, and quantum computation.
\end{IEEEbiographynophoto}

\begin{IEEEbiographynophoto}{Genwei Li}
is currently pursuing the B.S. degree in mathematics and applied mathematics with the School of Mathematical Sciences, Beihang University, Beijing, China. His academic interests include numerical linear algebra, scientific computing, and optimization algorithms.
\end{IEEEbiographynophoto}

\begin{IEEEbiographynophoto}{Delin Chu}
(Senior Member, IEEE) received the Ph.D. degree from the Department of Applied Mathematics, Tsinghua University, Beijing, China, in 1991. He is currently with the Department of Mathematics, National University of Singapore, Singapore. He is an Associate Editor of several journals, including Automatica, SIAM Journal on Scientific Computing, and SIAM Journal on Matrix Analysis and Applications. His research interests include data mining, numerical linear algebra, scientific computing, numerical analysis, and matrix theory and computations.
\end{IEEEbiographynophoto}

\end{document}